%% file: aanda.tex
\documentclass{aa}  
\usepackage{dblfloatfix}
\usepackage{graphicx}

\usepackage{natbib}
\usepackage{capt-of}
\usepackage{textcomp}
\usepackage{epstopdf}
\usepackage{rotating}
\usepackage{siunitx}
\usepackage{multirow}
\usepackage{hhline}
\usepackage{float}
\floatstyle{plaintop}
\restylefloat{table}
\usepackage{threeparttable}
\usepackage{longtable}
\usepackage{import}
\usepackage{array}
\usepackage{amsmath}
\usepackage{mwe}
\usepackage{subfig}
\usepackage[version=4]{mhchem}

\usepackage[colorlinks=true,allcolors=blue]{hyperref}

\usepackage{hyperref}
\begin{document} 

\renewcommand{\thefootnote}{\alph{footnote}}
\renewcommand{\thefootnote}{\fnsymbol{footnote}}

   \title{Infrared spectra of complex organic molecules in astronomically relevant ice mixtures}
   \subtitle{V. Methyl cyanide (acetonitrile)}
\titlerunning{}

   \author{M. G. Rachid\inst{1},    
           W. R. M. Rocha\inst{1}
          \and
            H. Linnartz
          \inst{1}          }

   \institute{Laboratory for Astrophysics, Leiden Observatory, Leiden University, P.O. Box 9513, NL 2300 RA Leiden, The Netherlands.\\
              \email{rachid@strw.leidenuniv.nl}
             }

   \date{Received ZZ; accepted YY}



  \abstract
   {The increasing sensitivity and resolution of ground-based telescopes have enabled the detection of gas-phase complex organic molecules (COMs) across a variety of environments. Many of the detected species are expected to form on the icy surface of interstellar grains and transfer later into the gas phase. Therefore, icy material is regarded as a primordial source of complex molecules in the interstellar medium. Upcoming  James Webb Space Telescope (JWST)  observations of interstellar ices in star-forming regions will reveal infrared (IR) features of frozen molecules with unprecedented resolution and sensitivity. To identify COM features in the JWST data, laboratory IR spectra of ices for conditions that simulate interstellar environments are needed.}
   {This work provides laboratory mid-IR spectra of methyl cyanide (CH$_3$CN, also known as acetonitrile) ice in its pure form and mixed with known interstellar molecules at cryogenic temperatures. The spectroscopic data presented in this work will support the interpretation of JWST ice observations and are made available to the community through the Leiden Ice Database for Astrochemistry (LIDA).}
   {Fourier transform IR spectroscopy is used to record the mid-IR spectra (500 - 4000 cm$^{-1}$/ 20 - 2.5 $\mu$m, with a resolution of 1 cm$^{-1}$) of methyl cyanide (acetonitrile, CH$_3$CN) mixed with H$_2$O, CO, CO$_2$, CH$_4$, NH$_3$, H$_2$O:CO$_2$, and H$_2$O:CH$_4$:NH$_3$, at temperatures ranging from 15 K - 150 K. The refractive index (at 632.8 nm) of pure amorphous CH$_3$CN ice at 15 K and the band strength of selected IR transitions are also measured.}
   {We present a variety of reference mid-IR spectra of frozen CH$_3$CN that can be compared to astronomical ice observations. The peak position and full width at half maximum (FWHM) of six absorption bands of frozen methyl cyanide in its pure form and mixed ices, at temperatures between 15 K - 150 K, are characterized. These bands are the following: the CH$_3$ symmetric stretching at 2940.9 cm$^{-1}$ (3.400 $\mu$m), the  CN stretching  at 2252.2 cm$^{-1}$ (4.440 $\mu$m), a peak resulting from a combination of different vibrational modes at 1448.3 cm$^{-1}$ (6.905 $\mu$m), the CH$_3$ antisymmetric deformation at 1410 cm$^{-1}$ (7.092 $\mu$m), the CH$_3$ symmetric deformation at 1374.5 cm$^{-1}$ (7.275 $\mu$m), and the  CH$_3$ rock vibration at 1041.6 cm$^{-1}$(9.600 $\mu$m). Additionally, the apparent band strength of these vibrational modes in mixed ices is derived. The laboratory spectra of CH$_3$CN are compared to observations of interstellar ices toward W33A and three low-mass Young Stellar Objects (YSO). Since an unambiguous identification of solid methyl cyanide toward these objects is not possible, upper limits for the CH$_3$CN column density are determined as $\leq$ 2.4 $\times$ 10$^{17}$ molecules cm$^{-2}$  for W33A and 5.2 $\times$ 10$^{16}$, 1.9 $\times$ 10$^{17}$, and 3.8 $\times$ 10$^{16}$ molecules cm$^{-2}$ for EC92, IRAS 03235, and L1455 IRS3, respectively. With respect to solid H$_2$O, these values correspond to relative abundances of 1.9, 3.1, 1.3, and 4.1 percent, for W33A, EC92, IRAS 03235, and L1455 IRS3, respectively.}
   {}


   \maketitle
%
\section{Introduction}

An increasing number of molecules has been detected in the interstellar medium (ISM) over the past decades, adding to more than 240 species \citep{mcguire2022}. Listed in this inventory are many organic molecules with more than six atoms, which in astrochemistry are referred to as complex organic molecules \citep[COMs,][]{herbst2009complex}. These COMs were identified across a diversity of environments, ranging from cold starless cores \citep[e.g.,][]{scibelli2020prevalence,Jorgensen2020} to objects in different phases of protostellar evolution \citep[e.g.,][]{remijan2005survey,herbst2009complex,jorgensen2016alma,van2020complex}, and extragalactic sources \citep[e.g.,][]{henkel1987detection,aladro2011lambda,sewilo2019complex}. Although observations and laboratory studies indicate that these species are mainly formed in interstellar ices \citep[e.g.,][]{Elsila2007, bisschop2007testing,requena2008galactic, Jimenez_Escobar2016, Chuang2020, Chuang2021, Ioppolo2021}, the only interstellar COM securely detected in the solid state to date is CH$_3$OH \citep{Skinner1992}. The lack of icy COM detections is mainly driven by instrumental limitations and spectral overlap of molecular absorption features. Observing missions with the {\it Infrared Space Observatory} (ISO) and the {\it Spitzer} Space Telescope allowed systematic observations of ices toward high- and low-mass star-forming regions \citep[e.g.,][]{Gibb2004,Boogert2008}. {\it Spitzer} also observed ice bands toward quiescent molecular clouds, thus enabling the study of the ice formation in starless cores \citep{Boogert2011}.  
So far, only the simple and more abundant molecular species have been detected in interstellar ices, such as H$_2$O, CO$_2$, CO, CH$_4$, NH$_3$, and CH$_3$OH  \citep{Oberg2011_spitzer}.  With the recent launch of the James Webb Space Telescope (JWST), this scenario is expected to change. With the high sensitivity, spatial and spectroscopic resolution of JWST, weak features of frozen molecules may be revealed. To support the upcoming JWST ice observations, laboratory studies aiming at characterizing COMs in relevant astronomical ices have been carried out during the past years (see for example \citealt{mate2017laboratory,urso2017infrared,luna2018densities,hudson2018thiol,van2018infrared,scire2019mid,palumbo2019laboratory,TvS2021, Rachid2020, Rachid2021,Muller2021,gerakines2022direct}). These works are of paramount importance for the correct assignment of features in interstellar ice spectra since the molecular vibrational bands are sensitive to ice composition, temperature, and morphology \citep[e.g.,][]{pontoppidan2008c2d,cuppen2011co}. Therefore,  interpreting astronomical ice observations using laboratory spectra of molecules in realistic astronomical conditions brings not only information about the ice chemical composition, but also the ice matrix and processes taking place in the solid phase in the ISM \citep{ehrenfreund1998ice}.

\begin{figure}
\centering
\includegraphics[width=0.55\linewidth]{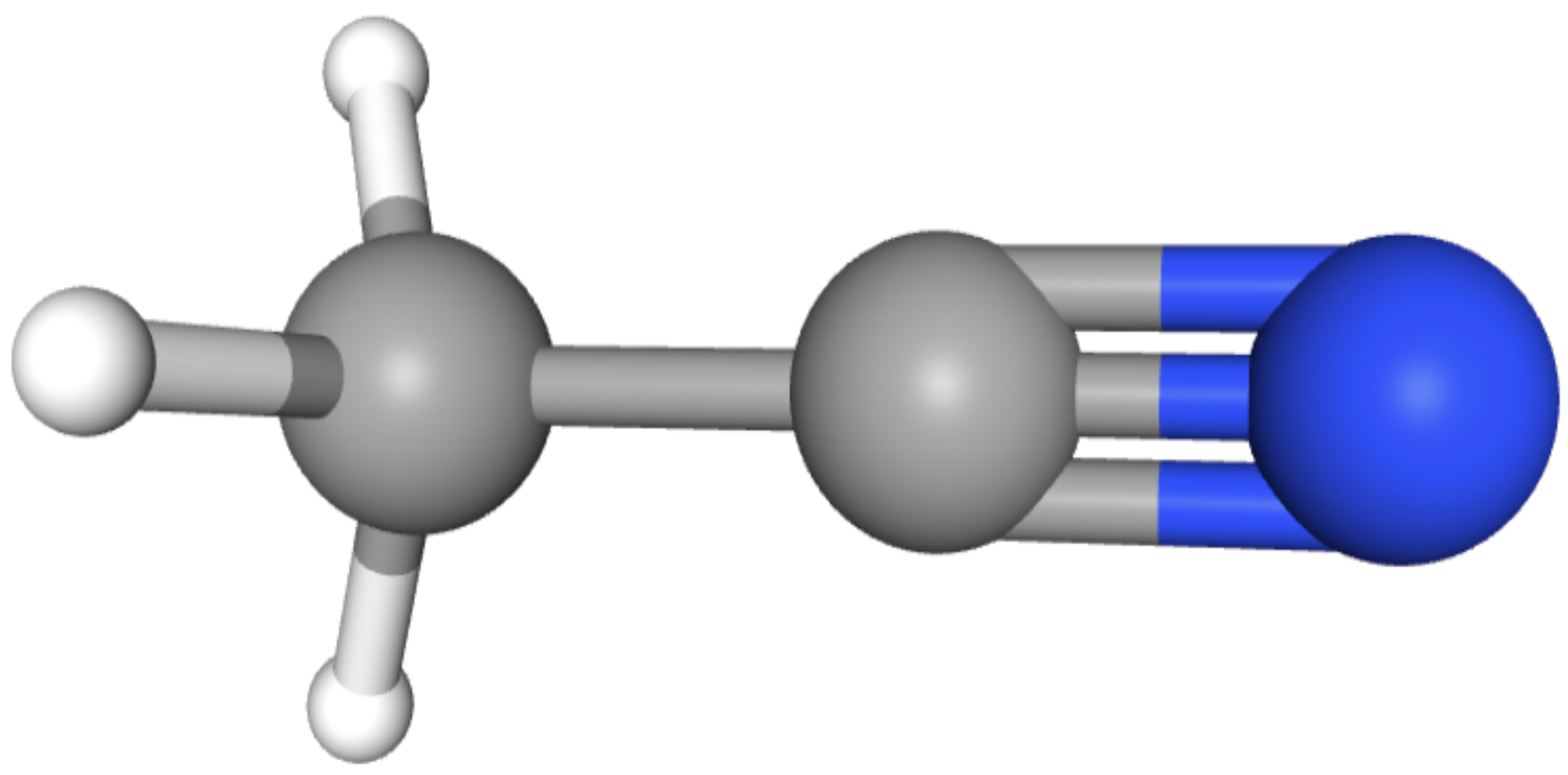}
\caption{Three-dimensional molecular structure of CH$_3$CN (hydrogen: white; carbon: gray; nitrogen: blue).}
\label{fig:ch3cn}
\end{figure}

This work is a study of the spectral features of methyl cyanide (or acetonitrile, CH$_3$CN) in astronomical relevant ices. Methyl cyanide (Figure \ref{fig:ch3cn}) is a symmetric top molecule, which makes it an important species to trace the different physical conditions in the ISM \citep{loren1984methyl}. Methyl cyanide has been identified across a variety of environments, such as in the galactic diffuse molecular gas \citep{liszt2018chemical}, in low and high mass star-forming regions \citep{araya2005ch3cn}, protoplanetary disks \citep{oberg2015comet,loomis2018,bergner2018survey}, photodissociation regions \citep{guzman2014chemical}, extragalactic sources \citep{mauersberger1991dense} and in our Solar System, in the atmosphere of Titan \citep{marten2002new}, and in the material  outgassed by comets  \citep{bockelee2000new,dutrey1996comet,goesmann2015organic}.

Despite its ubiquitous presence in the ISM, the interstellar formation of methyl cyanide remains a puzzle, and both gas and solid-state mechanisms have been proposed. In the gas phase, the radiative association of HCN and CH$_3$$^+$:\newline

\noindent \ce{HCN + CH3^+ $\rightarrow$ CH3CNH^+ + h\nu,}\newline

\noindent followed by dissociative recombination:\newline

\noindent \ce{CH3CNH^+ + e^- $\rightarrow$ CH3CN +H,} \newline

\noindent can yield CH$_3$CN \citep{huntress1979synthesis,willacy1993gas,mackay1999ch3cn}. In the solid phase, CH$_3$CN is expected to form through radical-radical reaction: \\

\noindent \ce{CH3^. + CN^. $\rightarrow$ CH3CN,}\newline

\noindent or through successive  hydrogenation of C$_2$N \citep{garrod2008complex}. Recently, \citet{volosatova2021direct} experimentally showed that CH$_3$CN can also be formed upon X-ray irradiation of CH$_4$:HCN in matrix isolated ices. To determine the dominant formation pathways in different astronomical conditions, more spatially resolved observations of CH$_3$CN and laboratory measurements of reaction rate constants are needed. Nevertheless, the gas-phase abundance of methyl cyanide observed in protoplanetary disks indicates that the formation of CH$_3$CN may happen in the ice phase \citep{oberg2015comet,loomis2018}. This assumption is in line with recent observations of methyl cyanide and other COMs in the disk around the young star V883 Ori by \citet{lee2019ice}. The quick increase in the temperature of V883 Ori's inner disk, caused by outbursts, sublimates the ice mantles covering the dust grains. Consequently, the content of the ice material is injected into the gas phase, providing a unique chance to probe the solid-phase molecular inventory through sensitive millimeter ALMA observations.

In interstellar ices, methyl cyanide can be a precursor of a range of N-bearing COMs. Experimental works about the UV photolysis of CH$_3$CN and CH$_3$CN-containing ices were shown to result in a variety of more complex species, among them amino acids, amines, and amides \citep{hudson2008amino,abdulgalil2013laboratory,bulak2021photolysis}. In a recent study, \cite{bulak2021photolysis} showed that the UV irradiation of CH$_3$CN in a H$_2$O environment has the potential to form several COMs containing both oxygen and nitrogen. Thus, CH$_3$CN is an important piece for unveiling the chemistry taking place in the ISM.

To identify CH$_3$CN ice signatures in future JWST data, the spectrum of this species in the solid state is needed. The absorption profile of a molecule (e.g., peak positions, bandwidths, and intensities) is sensitive to the ice composition, morphology, and temperature. Therefore, the spectrum of  CH$_3$CN  in simulated interstellar ice conditions is necessary. The mid-IR spectra of pure CH$_3$CN ice in its amorphous and different crystalline forms have been reported in the literature (see, for example, \citet{russo1996laboratory,hudson2004reactions,abdulgalil2013laboratory,hudson2020preparation}). The spectra of CH$_3$CN mixed with H$_2$O, CH$_3$OH, and CH$_3$CH$_2$OH in a range of different temperatures have been also investigated \citep{bhuin2015interaction,methikkalam2017interaction}. These studies report changes in the infrared profile of methyl cyanide in different ice mixtures and explore the molecular interactions taking place in the solid phase (e.g., hydrogen bonding). To simulate more realistic interstellar ices, it is necessary to study samples containing known interstellar ice molecules and low concentrations of CH$_3$CN. Also important is to report the changes in peak position, bandwidth, and strength of CH$_3$CN features in different ice mixtures to offer astronomers a quick source of spectroscopic parameters.

This work investigates the spectra of CH$_3$CN mixed with H$_2$O, CO, CO$_2$, CH$_4$, and NH$_3$, at temperatures ranging from 15 K - 150 K. The position and FWHM of selected CH$_3$CN bands are characterized and their potential to identify methyl cyanide in astronomical ice observations is discussed. This paper is structured as follows. Section 2 brings the experimental setup and measurement protocols. Section 3 reports the changes in the profile of the selected methyl cyanide bands in different ice mixtures and temperatures. Section 4 explores the potential of methyl cyanide features to trace this species in ice observations and shows the use of the laboratory spectra to interpret astronomical data. Section 5 summarizes the conclusions of the work. The appendices bring the infrared profile, peak position, FWHM, and the integrated absorbance of the characterized methyl cyanide bands in the different ice mixtures and temperatures. The complete data set recorded in this work is available from the Leiden Ice Database for Astrochemistry\footnote{\url{https://icedb.strw.leidenuniv.nl/}} (LIDA, Rocha et al. submitted.).

\section{Methodology}

All the measurements are performed using IRASIS (InfraRed Absorption Setup for Ice Spectroscopy), a vacuum setup (base pressure $\sim$ 2 $\times$ 10$^{-9}$ mbar) dedicated to acquiring mid-IR spectra and visible refractive index (at 632.8 nm) of ice samples.  The analyzed ices are grown by background deposition of a gaseous sample onto a germanium (Ge) substrate kept at 15 K by a closed-cycle helium cryostat. Typically, the ices are grown for 10 $-$ 20 minutes, which results in a thickness of around 1 $-$ 1.5 $\mu$m on each side of the germanium substrate. After the ice deposition, a Fourier transform infrared spectrometer is used to record the mid-infrared spectrum (20 $-$ 2.5 $\mu$m, 500 - 4000 cm$^{-1}$) in transmission mode at resolution 1.0 cm$^{-1}$ and averaging 128 scans. Subsequently, the sample is heated at a rate of 25 K/hour until its complete desorption. The spectrum of the ice sample is recorded during the heating and it takes around 2 minutes to obtain the complete spectrum. This time interval results in an uncertainty of 1 K in the temperature associated with the spectrum. For all the ice mixtures, a single sample is prepared and studied. For the pure CH$_3$CN ice, at least three samples are prepared, and since the IR profile of methyl cyanide bands in all the samples is identical, the spectrum of one of them is chosen to be reported. For a detailed description of the setup and methodology see \citet{Rachid2021}. 

The changes in the IR profile of methyl cyanide when mixed with different and abundant interstellar ice molecules (H$_2$O, CO, CO$_2$, CH$_4$, and NH$_3$) are investigated. For this purpose, binary mixtures containing methyl cyanide are prepared at ratios 1:5, 1:10, and 1:20 (i.e., for each CH$_3$CN molecule in the mixture, there are 5, 10, or 20 molecules of the second component). A three and a four components mixtures are also analyzed: CH$_3$CN:H$_2$O:CH$_4$:NH$_3$(1:20:2:2) and CH$_3$CN:H$_2$O:CO$_2$(1:5:2). These more complex samples have a composition that is closer to "real" interstellar ices. The gaseous samples are prepared in a separate mixing system by the sequential addition of vapor or gas components to a glass bulb. The liquid components (H$_2$O and CH$_3$CN) are purified by at least three freeze-thaw cycles before preparing the mixture. The gaseous mixture in the bulb has a total pressure of 20 mbar, and the partial pressure of each molecular component corresponds to their fraction in the mixture. It is assumed in this work that the ratio of a molecular species in the final ice sample is the same as in the gaseous mixture. The estimated error in the mixture ratios is estimated as 20, 24, and 30  percent for the two, three, and four-component mixtures, respectively \citep{Rachid2021}. The gases and liquids used for preparing the gaseous mixtures are methyl cyanide (acetonitrile, VWR, $\leq$ 10 ppm of water), water (Milli-Q, Type I), carbon monoxide (Linde gas, 99.997\%), carbon dioxide (Linde Gas, 99.95\%) methane (Linde Gas 99.995\%), and ammonia (Sigma Aldrich, anhydrous $\geq$ 99.99 \%).

\subsection{Refractive index and ice thickness}

The real part of the refractive index (n) of a solid sample can be measured through a technique that consists in acquiring the interference pattern of two individual laser beams, hitting the growing ice at the same spot but under two different angles \citep{romanescu2010refractive,beltran2015double}.
The real refractive index is related to the period of the generated interference patterns by: 

\begin{equation}
    n = \sqrt{\frac{\sin^2 \theta_0-\gamma^2 \sin^2 \theta_1}{1 - \gamma^2}},
\label{refractive}
\end{equation}
where  $\gamma = \frac{T_1}{T_0}$ is the ratio between the periods of the interference patterns, T$_1$ and T$_0$,  produced by the laser beams hitting the substrate at angles $\theta_1$ and $\theta_0$, respectively. In this work,  a red HeNe laser ($\lambda$ = 632.8 nm) hitting the ice at angles of  $\theta_0$ = 8.\ang{0} $\pm$ 0.\ang{5} and $\theta_1$ =  50.\ang{0} $\pm$ 0.\ang{5} is employed. For this wavelength, the measured refractive index of pure CH$_3$CN ice at 15 K is n = 1.28 $\pm$ 0.02. The uncertainty in this value is calculated from the standard deviation of five independent measurements. This value is slightly lower than values found in the literature for amorphous CH$_3$CN at 15 K, \citep[n = 1.334,][]{hudson2020preparation} and at 30 K \citep[n = 1.31,][]{2010ApJS96M}.

Once the refractive index of the ice is known, its thickness can be measured by recording a single laser interference measurement during the ice deposition. The ice thickness is related to the number of fringes (m) in an interference pattern by:

\begin{equation}
    d = \cfrac{m \lambda}{2 \sqrt{n^2 - sin^2 \theta}} , 
\label{thickness}    
\end{equation}
where  $\lambda$ and $\theta$ are the wavelength and incidence angle of the laser light, respectively, and $n$ is the ice's refractive index. The thickness of the pure methyl cyanide ice analyzed in this work is around 2.63 $\mu$m (considering both sides of the substrate). 

\subsection{IR band strengths}

To derive the band strength of an IR absorption feature (\textit{A}, in units cm molec$^{-1}$), it is necessary to know its integrated absorption as well as the number of absorbing species in the pathway of the probing IR beam, given by the column density (N, in units of molecules cm$^{-2}$, \citealt{d1986time}). The column density of a species in pure ice is related to the ice thickness and density (assuming that the thickness of the ice is homogeneous in the area probed by the beam) by:

\begin{equation}
  N = \frac{d \;\rho \;N_A}{M} ,
\label{column}
\end{equation}
where d and $\rho$  are the ice thickness (in cm) and density (in g cm$^{-3}$), M is the molar mass (in g mol$^{-1}$) of the molecule being deposited, and N$_A$ is Avogadro's number. This expression can be combined with the Lambert-Beer law to provide the band strength in terms of the ice parameters and integrated absorbance of an absorption band:

\begin{equation}
  A = \frac{2.3\;M\;\alpha}{\rho\; \; N_A \; \beta} ,
    \label{bs}
\end{equation}
where M = 41.05 g mol$^{-1}$ is the molecular mass of methyl cyanide, $\rho$ = 0.778 g cm$^{-3}$ is the ice density, from \citet{hudson2020preparation}, $\alpha$ is the growth rate of the integrated absorbance (cm$^{-1}$ s$^{-1}$), and $\beta$ is the ice thickness growth (cm s$^{-1}$). In Equation \ref{bs}, the growth of the ice thickness and integrated absorbance during the ice deposition are used instead of single measurements of thickness and integrated absorbance at a given time. This procedure is valid as long as both growth rates are linear in time, which is the case in our measurements. Using equation \ref{bs}, the refractive index of CH$_3$CN at 15 K and the measured growth rate of the ice thickness and integrated absorbance of the CH$_3$CN features, the band strengths for pure CH$_3$CN ice at 15 K are obtained (Table \ref{band-strength}). The uncertainty in the values listed in Table \ref{band-strength} is calculated through error propagation of the terms in Equation \ref{bs}. The uncertainties in  $\alpha$ and $\beta$, follow from the uncertainty in the linear fit of the thickness (or integrated absorbance) with time. These uncertainties amount to a maximum of 4 percent for the ice thickness growth and 10 percent for the integrated absorbance. For the density value, an uncertainty of 10 percent is taken to avoid an underestimation of the uncertainty. The final uncertainty in the band strengths resulting from these assumptions is smaller than 15 percent. These band strengths derived in this work are generally in good agreement with most of the values measured by \citet{d1986time}, also shown in Table \ref{band-strength}, and the CN stretching band strength value derived by \citet{hudson2004reactions}, A = 2.2 $\times$ 10$^{-18}$ cm molecule$^{-1}$. Small differences in the values derived in previous works can be attributed to variations in the experimental conditions, integration boundaries in the IR spectrum, and baseline correction. The band strength values derived in this work are employed to calibrate the deposition rate of methyl cyanide in the measurements.

\begin{table*}[h]
\caption{IR band strength of amorphous CH$_3$CN ice at 15 K derived in this work.}
\begin{center}
\begin{tabular}{c  c  c  c c}
\hline
         Peak position & Integrated region & Assignment  & A & A$^{**}$ \\
         (cm${}^{-1}$) & (cm${}^{-1}$) &  & \multicolumn{2}{c}{ (10${}^{-18}$ cm molec${}^{-1}$) }  \\
        \hline 
         3001.7 & 3030 - 2975  & CH$_3$ antisymmetric stretching & 1.5 $\pm$ 0.2 & 1.5\\
         2940.9 &2955 - 2923 & CH$_3$ symmetric stretching & 0.53 $\pm$ 0.08& 0.47 \\
         2289 &2310 - 2270 & Combination of modes & 0.62 $\pm$ 0.09 &\multirow{2}{*}{2.3}\\
         2252.2 &2265 - 2235 & CN stretching & 1.9 $\pm$ 0.3& \\
         
         1448.3 &1490 - 1435 & Combination of modes & 2.9 $\pm$ 0.4& 2.2\\
         1410.0 &1434 - 1398.5 & CH$_3$ antisymmetric def. & 1.9  $\pm$ 0.3& 1.4 \\
         1374.5 & 1395 - 1357 & CH$_3$ symmetric def. & 1.2  $\pm$ 0.2 & 1.0\\

         1041.6 &1100 - 1025 &CH$_3$ rock & 1.6 $\pm$ 0.2& 1.3\\

         919.6 & 940 - 913.5 & C-C stretching & 0.35 $\pm$ 0.05& 0.2\\

        \hline
\end{tabular}

\begin{tablenotes}
    \item[\emph{}]{$^*$Band assignments following \citealt{milligan1962solid,russo1996laboratory}, and references therein. \\}

    \item[\emph{}]{$^{**}$ Values from 
    \citealt{d1986time}. Estimated uncertainties are around 20\%.}

\end{tablenotes}
\end{center}

\label{band-strength}
\end{table*}

In the ice mixtures, the apparent band strength of a feature \emph{i} (A$_i'$) is derived using the column density of methyl cyanide in the mixture (N$_{mix}$) and the integrated absorbance of the feature by:

\begin{equation}
    A{_i'} = 2.3 \; \frac{\int_i abs(\nu) \ d\nu}{N_{mix}},
\end{equation}
where $\int_i abs(\nu) \ d\nu$ is the integrated absorbance of the feature \emph{i}. We note that the absorbance is measured on the logarithm base 10 scale and is related to the transmittance (T) of light through the sample by Abs = -log(T). In this work, the column density and deposition rate of methyl cyanide in the ice mixtures (N$_{mixt}$) are derived in the following way. First, the deposition rate of pure CH$_3$CN is determined when the leak valve is kept at a fixed position and a bulb filled with 20 mbar of methyl cyanide vapor is connected to the dosing line. This procedure was repeated three times and the average value obtained for the deposition rate of pure CH$_3$CN was around 4.7 $\times$ 10$^{15}$ molecules  cm$^{-2}$ s$^{-1}$. Next, this value is used to derive the CH$_3$CN deposition rate when a gas mixture containing a total pressure (adding all components) of 20 mbar is employed. This is done by multiplying  the deposition rate for the pure CH$_3$CN by its fraction in the gaseous mixtures: 
\begin{equation}
    D_{CH_3CN,mixt} = D_{CH_3CN,pure} \times f,
\end{equation}
where D$_{CH_3CN,mixt}$ and  D$_{CH_3CN,pure}$  are the deposition rate of CH$_3$CN when depositing an ice mixture and pure CH$_3$CN, respectively, and $f$ is the fraction of CH$_3$CN in the gaseous mixture. As an example, for mixtures in which CH$_3$CN is diluted in a 1:10 ratio, the deposition rate of CH$_3$CN would be D$_{CH_3CN,mixt(1:10)}$ = 4.7 $\times$ 10$^{15}$ $\times$ $\frac{1}{11}$ = 4.3 $\times$ 10$^{14}$ molecules cm$^{-2}$ s$^{-1}$. Here, we assume that the deposition of methyl cyanide is not altered differently by the different gas components. This methodology comes with rather large uncertainties in the mixing ratio, which are estimated to amount to up to 30 percent.

The relative band strengths of methyl cyanide features in ice mixtures (i.e., the apparent band strength normalized with respect to the value for the pure ice) are calculated through the relation:

\begin{equation}
    \eta = \frac{A{_i'}}{A_i},
    \label{ba}
\end{equation}
where A$_i$ is the band strength of the feature in pure CH$_3$CN ice at 15 K derived in this work. The relative band strengths are derived for all the characterized bands in ices at 15 K. In conclusion, it is important to stress that currently it is not possible to reduce the large uncertainties in relative band strengths, which can be as high as 40 percent.

\section{Results}
\label{results}

\begin{figure*}[ht]
\includegraphics[width=\hsize]{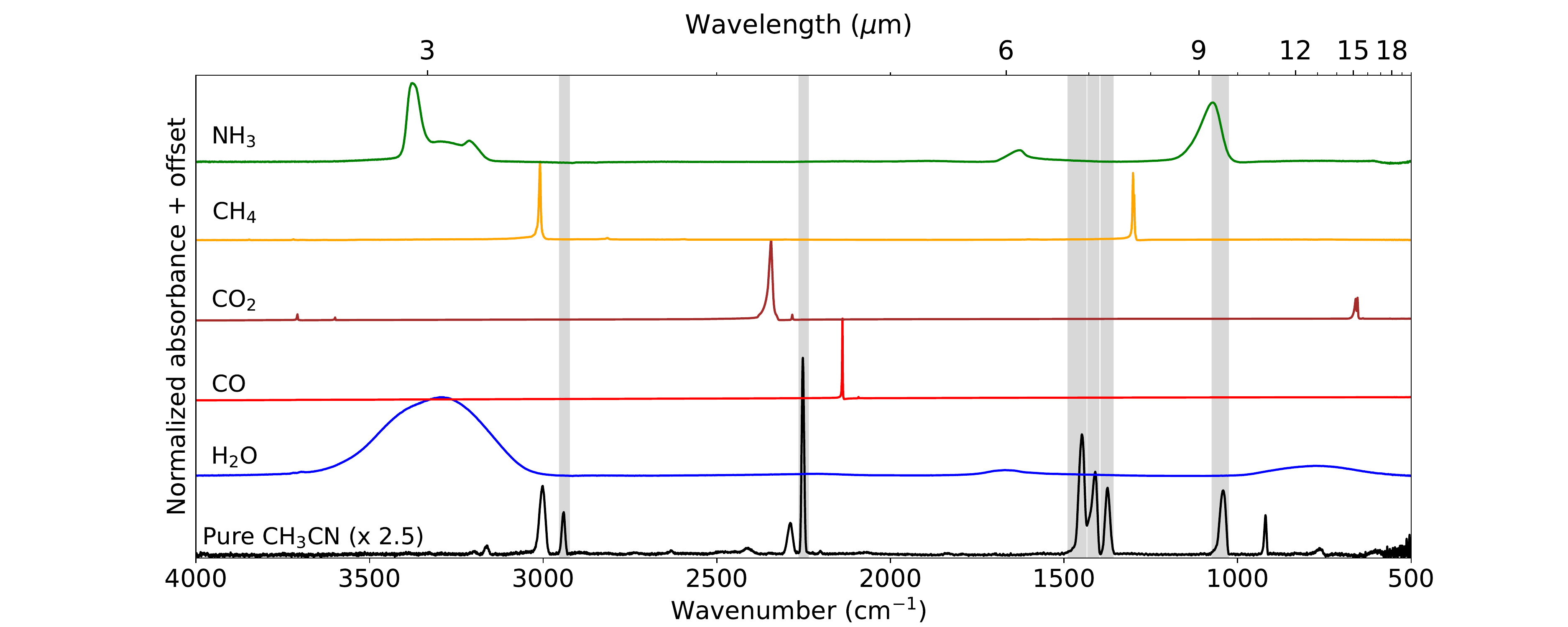}
\caption{Infrared spectra of pure methyl cyanide (CH$_3$CN) ice and the different molecules present in the ice mixtures analyzed in this work (H$_2$O, CO, CO$_2$, CH$_4$, and NH$_3$), all at 15 K.  The methyl cyanide bands characterized in this work are highlighted by a shaded area.}
\label{all_molecules}
\end{figure*}

This section describes the changes in the IR profile of selected methyl cyanide absorption features in the pure form and mixed with known interstellar ice components.  Figure \ref{all_molecules} shows the IR spectra of pure CH$_3$CN ice and the other matrix molecules used in this work: H$_2$O, CO, CO$_2$, CH$_4$, and NH$_3$. Despite the relatively high deposition rates, the solid samples deposited at 15 K are amorphous. This can be concluded from the peak position and broad profile of the IR peaks of the pure methyl cyanide ices, characteristic of its amorphous structure (see \citealt{hudson2004reactions,abdulgalil2013laboratory,bhuin2015interaction,hudson2020preparation,smith2021crystallization}). The crystalline forms of methyl cyanide display a rather different peak profile in the 1500  - 1000 cm$^{-1}$ region, in which the bands assigned to the CH$_3$ symmetric and antisymmetric deformation modes and the CH$_3$ rock display multiple components.  In the measurements presented here, the peaks characteristics of crystalline CH$_3$CN are only observed after heating the pure ice to 100 K or temperatures higher than 120 K for mixed ices, as discussed in the following sections.  Except for the CO-containing ices, the major components of the mixed ices used in this work are deposited as amorphous ice. This is evident by the comparison of their IR spectra with available literature data (see, for example, \citealt{bouilloud2015bibliographic} for the profile of amorphous H$_2$O and the other molecules used in the mixture, \citealt{gerakines2015infrared} for the profile of amorphous CH$_4$ ice, \citep[see][]{ehrenfreund1999laboratory,palumbo2000infrared} for CO$_2$; \citealt{zanchet2013optical} for the NH$_3$ profile, and \citealt{ehrenfreund1997} for the pure CO profile).
In the CO-containing ices, CO is likely deposited in a polycrystalline form, as the formation of amorphous CO ice is only possible at temperatures below 8 K \citep{he2022radical}.

The  CH$_3$CN absorption bands highlighted by the shaded areas in Figure \ref{all_molecules} are selected for peak position and FWHM characterization. These are chosen according to their strength and marginal or no overlap with spectral features of other abundant species. The selected bands are the following: the CH$_3$ symmetric stretching at 2940.9 cm$^{-1}$ (3.400 $\mu$m), the  CN stretching  at 2252.2 cm$^{-1}$ (4.440 $\mu$m), a combination of modes at 1448.3 cm$^{-1}$ (6.905 $\mu$m), the CH$_3$ antisymmetric deformation at 1410 cm$^{-1}$ (7.092 $\mu$m), the CH$_3$ symmetric deformation at 1374.5 cm$^{-1}$ (7.275 $\mu$m), and the  CH$_3$ rock at 1041.6 cm$^{-1}$(9.600 $\mu$m). The potential of these bands as tracers of  CH$_3$CN in interstellar ice observations is discussed in Section \ref{astrochem}. The peak positions, assignments and band strengths measured for the amorphous CH$_3$CN  ice at 15 K are listed in Table \ref{band-strength}.

The infrared profile of methyl cyanide strongest mid-IR band, the CN stretching, in different ice mixtures and temperatures is shown in Figure \ref{fig:2252}. Similarly, the profiles of the other bands characterized in this work are displayed in figures \ref{fig:2900}, \ref{fig:1450}, \ref{fig:1400}, \ref{fig:1350}, and \ref{fig:1040} in Appendix A. In these figures, binary mixtures in a ratio of 1:10 are used to illustrate the different band shapes of methyl cyanide absorption features in the mixtures. The discussion on the peak position and FWHM changes in the binary mixtures throughout this section is also based on the 1:10 mixtures, since the 1:5 and 1:20 binary mixtures present similar characteristics. The complete peak position and FWHM characterization for all the ice mixtures at all ratios and temperatures are available in the Tables from Appendix B. The integrated absorbance of the CH$_3$CN features (i.e., the band areas) in all ice mixtures at all temperatures is listed in Appendix C. These values are used for deriving the strength of the methyl cyanide bands in the mixtures at temperatures above 15 K. This procedure is described in details in Appendix C.

\subsection{The 4000 - 2500 cm$^{-1}$ region (2.5 - 4.0 $\mu$m)}

In the 4000 - 2500 cm$^{-1}$ region, solid amorphous CH$_3$CN presents two normal modes, assigned to the C-H antisymmetric stretching, around 3001.7 cm$^{-1}$  (3.331 $\mu$m), and the  C-H symmetric stretching, at 2940.9 cm$^{-1}$ (3.400 $\mu$m). A couple of weaker features associated to overtones are also observed in this region \citep{bec2019overtones} but will not be discussed here. At temperatures below 100 K, the peak position of the two CH$_3$CN normal modes does not change appreciably. Just above 100 K, there is an abrupt sharpening of both features, characteristic of a phase transition. These bands appear in the same region of the C-H features as other interstellar ice molecules. As can be seen in Figure \ref{all_molecules}, the C-H antisymmetric feature of methyl cyanide, at 3001.7 cm$^{-1}$, overlaps with the strong methane feature at 3010 cm$^{-1}$ (3.322 $\mu$m). The feature of the C-H symmetric stretching does not overlap with any of the ice components used in this work, which makes it an interesting band for further study. However, this peak overlaps with the C-H vibrational modes from other COMs, which compromises its potential in tracing methyl cyanide in interstellar ices, as discussed in Section \ref{astrochem}.

Figure \ref{fig:2900} shows the IR profile of the CH$_3$ symmetric stretching feature in ice mixtures. The FWHM versus peak position panel shows that for most ice mixtures this feature peaks between 2941 - 2946 cm$^{-1}$. Peak positions out of this range are found for CH$_3$CN:NH$_3$, in which this band shifts to low wavenumbers (2931.8  cm$^{-1}$) and shows a FWHM of 17.9 cm$^{-1}$ (the value for the pure ice is 10.9 cm$^{-1}$) and the CH$_3$CN:CO$_2$, in which this band shifts to high wavenumbers (around 2954.4  cm$^{-1}$). These differences in peak position and width can be evidenced in the bottom left panel of Figure \ref{fig:2900}, which displays the FWHM and peak position for this peak in the different ice mixtures and the range of temperatures studied here.

\subsection{The CN stretching feature - 2252 cm$^{-1}$}

The CN stretching vibration gives rise to an absorption band around 2252.2 cm$^{-1}$, which is the strongest methyl cyanide feature in the mid-IR.  By warming the pure ice to 95 -100 K, a pronounced narrowing and an increasing in the intensity of this band are observed. At temperatures above 120 K, this feature peaks around 2251 cm$^{-1}$ and has a FWHM of 1.3 cm$^{-1}$.

\begin{figure*}
\includegraphics[width=1.0\linewidth]{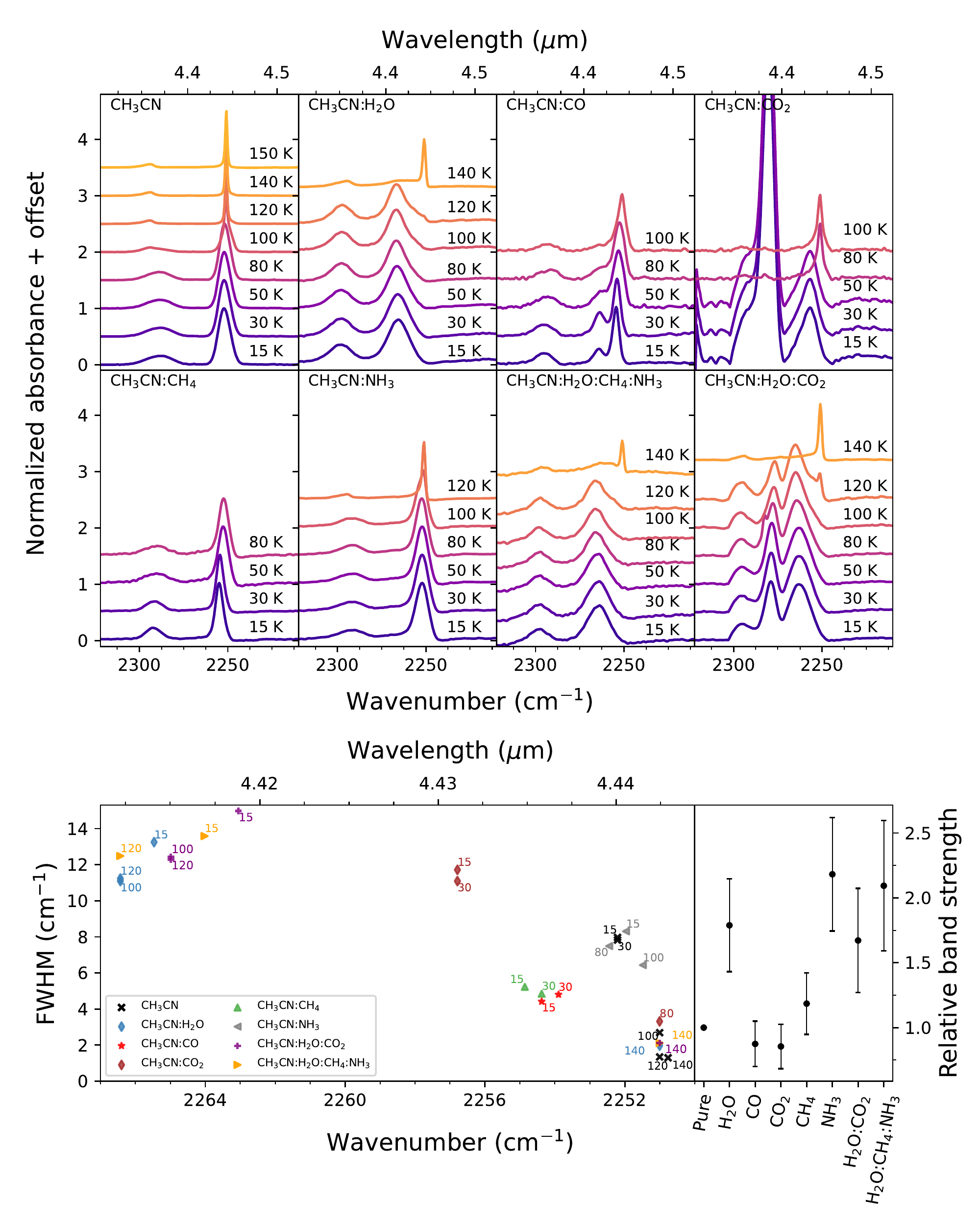}
\caption{Upper panel: Absorption profile of the CN stretching mode of methyl cyanide, around 2252.2 cm$^{-1}$ (4.440 $\mu$m), in pure and mixed ices. The ice spectra at different temperatures are indicated by different colors and labels. Bottom left: Peak position and FWHM of the CN stretching mode of CH$_3$CN in different ice mixtures at selected temperatures. Bottom right: relative band strengths  of the CN stretching band in different ice mixtures at 15 K.}
\label{fig:2252}
\end{figure*}

Figure \ref{fig:2252} shows the profile of the CN stretching band in the different ice matrices. In CH$_3$CN:CH$_4$ and CH$_3$CN:NH$_3$ ices, this feature peaks between 2257 and 2250 cm$^{-1}$ and has a profile similar to the pure methyl cyanide. In mixtures containing H$_2$O, CO$_2$, or CO, this feature shows some different characteristics.

In all the H$_2$O-containing ices, the CN stretching peaks at high wavenumbers, between 2263 - 2265 cm$^{-1}$, and it has a broader profile when compared to the other ice mixtures (FWHM $\sim$ 13 - 15 cm$^{-1}$). The different profile of this band in H$_2$O-rich ices can be seen in the panel displaying the  FWHM versus peak position plot (left bottom of Figure \ref{fig:2252}), where the points representing the water-rich ices with temperatures below 120 K appear grouped in the left top region of the graph. Above 125 K, the CN feature sharpens and shifts to 2251 cm$^{-1}$ for all the ice mixtures, as can be seen in the right corner of the panel. In the CH$_3$CN:CO$_2$ ice, the CN stretching band peaks at 2256.8 cm$^{-1}$ and has a broader profile than the band seen in the pure ice. Figure \ref{fig:2252} shows that the CN feature peaks near the $^{13}$CO$_2$ feature, around 2282 cm$^{-1}$. The $^{13}$CO$_2$ feature is seen in the panels showing  CO$_2$-containing mixtures but is still clearly separated from the CN stretching transition. In CH$_3$CN:CO ice, the CN stretching band peaks at 2253.9 cm$^{-1}$ and shows a small shoulder at high wavenumbers, which is not seen in other mixtures.

Interestingly, the peak position and FWHM of the CN stretching peak in the ice mixtures above 125 K are close to the values measured for the pure methyl cyanide at the same temperature. This is noticeable from the FWHM versus peak position panel in figure \ref{fig:2252}, in which the points representing the mixtures at high temperatures are grouped in the right corner of the plot. The similarities in the absorption profile of  CH$_3$CN-containing mixtures with pure CH$_3$CN at high temperatures are also observed for other bands and indicate that segregation of CH$_3$CN may take place in the ice mixtures.

\subsection{The 2000 - 500 cm$^{-1}$ region (5 - 20 $\mu$m)}

Pure amorphous methyl cyanide ice at 15 K has three strong absorption features in the region between 2000 - 1300 cm$^{-1}$, a strong feature at 1448.3 cm$^{-1}$ assigned to a combination of the CH$_3$ rock and CCN bending modes \citep{russo1996laboratory}, the CH$_3$ antisymmetric deformation at 1410 cm$^{-1}$ and the CH$_3$ symmetric deformation band, at 1374.5 cm$^{-1}$. In pure CH$_3$CN at temperatures above 100 K, these bands split into multiple components. Figures \ref{fig:1450}, \ref{fig:1400} and \ref{fig:1350} show the thermal evolution of these features in pure and mixed ices. The peak positions of the multiple features at high temperatures are displayed in Tables \ref{1450-mixtures}, \ref{1400-mixtures}, and \ref{1370-mixtures}.
 
In all the ice mixtures at 15 K, the peak position of the three features shows little changes, shifting less than 5 cm$^{-1}$ with relation to the pure ice.  As a general trend, the FWHM of these bands at low temperatures is narrower for the binary mixtures with CO, CO$_2$, and CH$_4$.  Above 100 K (and 120 K for H$_2$O-containing ices) these peaks sharpen and split, showing a profile similar to the pure crystalline ice in all the mixtures. At these high temperatures, the feature assigned to a combination of modes, near 1448 cm$^{-1}$,  shifts to around 1453 - 1455 cm$^{-1}$ for all the ice mixtures. The CH$_3$ antisymmetric deformation band, near 1410 cm$^{-1}$ in pure CH$_3$CN, splits into three components that peak around 1409, 1416, and 1419 cm$^{-1}$, similar to the features observed in pure methyl cyanide ice above 100 K. The band assigned to the CH$_3$ symmetric deformation band, near 1374.5 cm$^{-1}$, also splits into two components, around 1371.5 and 1378 cm$^{-1}$.

Below 1300 cm$^{-1}$, the pure CH$_3$CN ice has two strong peaks, at 1041.6 cm$^{-1}$ (CH$_3$ rock) and 919.6 cm$^{-1}$ (C-C stretching). The CH$_3$ rock feature overlaps with the low wavenumber ring of the NH$_3$ umbrella mode, but it can still be seen as a shoulder, even in diluted mixtures. Figure \ref{fig:1040} shows the profile of this band in different ice mixtures. The matrix components are subtracted from mixtures containing H$_2$O and NH$_3$ for better visualization of the CH$_3$ rock feature. The peak position of this ice feature does not change substantially in the ice mixtures, peaking between 1042 - 1037 cm$^{-1}$ in the ice mixtures at 15 K. The width of the CH$_3$ rock feature is larger in pure methyl cyanide than in the ice mixtures, as can be seen in the peak position versus FWHM plot in Figure \ref{fig:1040} and the values in Table \ref{1040-mixtures}. At temperatures above 120 K, the CH$_3$ rock feature splits into three narrower features in the ice mixtures, that peak around 1048, 1040, and 1036 cm$^{-1}$.

The C-C stretching appears as a weak feature in diluted mixtures. This peak has some further disadvantages for being used as a methyl cyanide tracer: it appears at very close wavelengths to the OH bending mode of formic acid, a species that has been tentatively detected toward some Young Stellar Objects \cite[YSOs;][]{schutte19966,schutte1999weak,Bisschop2007}. In an astronomical ice spectrum, this peak may be visible when methyl cyanide is present at relatively high column densities. At low methyl cyanide column densities, this peak will likely be overshadowed by the red wing of the H$_2$O libration mode (centered around 800 cm$^{-1}$) and the silicate feature around 10 $\mu$m. Because of these reasons, this band is not further characterized in this work.

\section{Astronomical implications}
\label{astrochem}

The spectroscopic data recorded in this work is an essential tool for identifying methyl cyanide features in observations of ices toward dense clouds and protoplanetary disks. Future JWST observations made at unprecedented sensitivity and spectral resolution \citep[e.g., R $\sim$ 3000 around 5 - 10 $\mu$m with the Mid InfraRed Instrument Medium Resolution Spectrometer - MIRI MRS,][]{labiano2021} will look for weak features of frozen COMs that are inaccessible from past ice surveys \citep{complicated}. Robust detection of new ice species will only be possible by the identification of multiple features using laboratory spectra as reference. Below, the potential of the methyl cyanide bands as tracers of this molecule in astronomical ice data is discussed.

\underline{Range between 2.5 and 3.4~$\mu$m}: The C-H stretching features of methyl cyanide, around 3.333 $\mu$m (3000 cm$^{-1}$) and 3.400 $\mu$m (2940.9~cm$^{-1}$), overlap with the low wavenumber (high wavelength) wing of the O-H stretching band of solid H$_2$O, around 3 $\mu$m. In combination with CH$_3$CN bands observed in other parts of the IR spectrum, the methyl cyanide features in this region can provide evidence for its presence in ice observations. However, assignments in this region should be done with care, due to the C$-$H stretching vibration bands of hydrocarbons and other COMs in this region (see, for example, \citealt{sandford1991interstellar,boudin1998constraints,boogert2004methane,Rachid2021}). 

\underline{Range between 4.3 and 4.5~$\mu$m}: The CN stretching feature, around 4.440~$\mu$m (2252.2~cm$^{-1}$), is methyl cyanide's strongest mid-IR band, and the one that offers the most potential for identifying this species in interstellar ice observations. This band does not overlap with features from abundant ice molecules or features from smaller CN-containing species. For example, the CN stretching transitions from HCN, CN$^-$ and OCN$^-$, peak around 4.76~$\mu$m, 4.78~$\mu$m, and 4.61~$\mu$m, respectively \citep{moore2003infrared,gerakines2004ultraviolet,van2004quantitative}. However, we highlight that the presence of strong features near 4.4 $\mu$m (e.g., from $^{13}$CO$_2$ around 4.38~$\mu$m and  OCN$^-$ between 4.59 - 4.61~$\mu$m) may overshadow the CN stretching feature if CH$_3$CN is present at a low column density.

\begin{figure}
\centering
\includegraphics[width=1\linewidth]{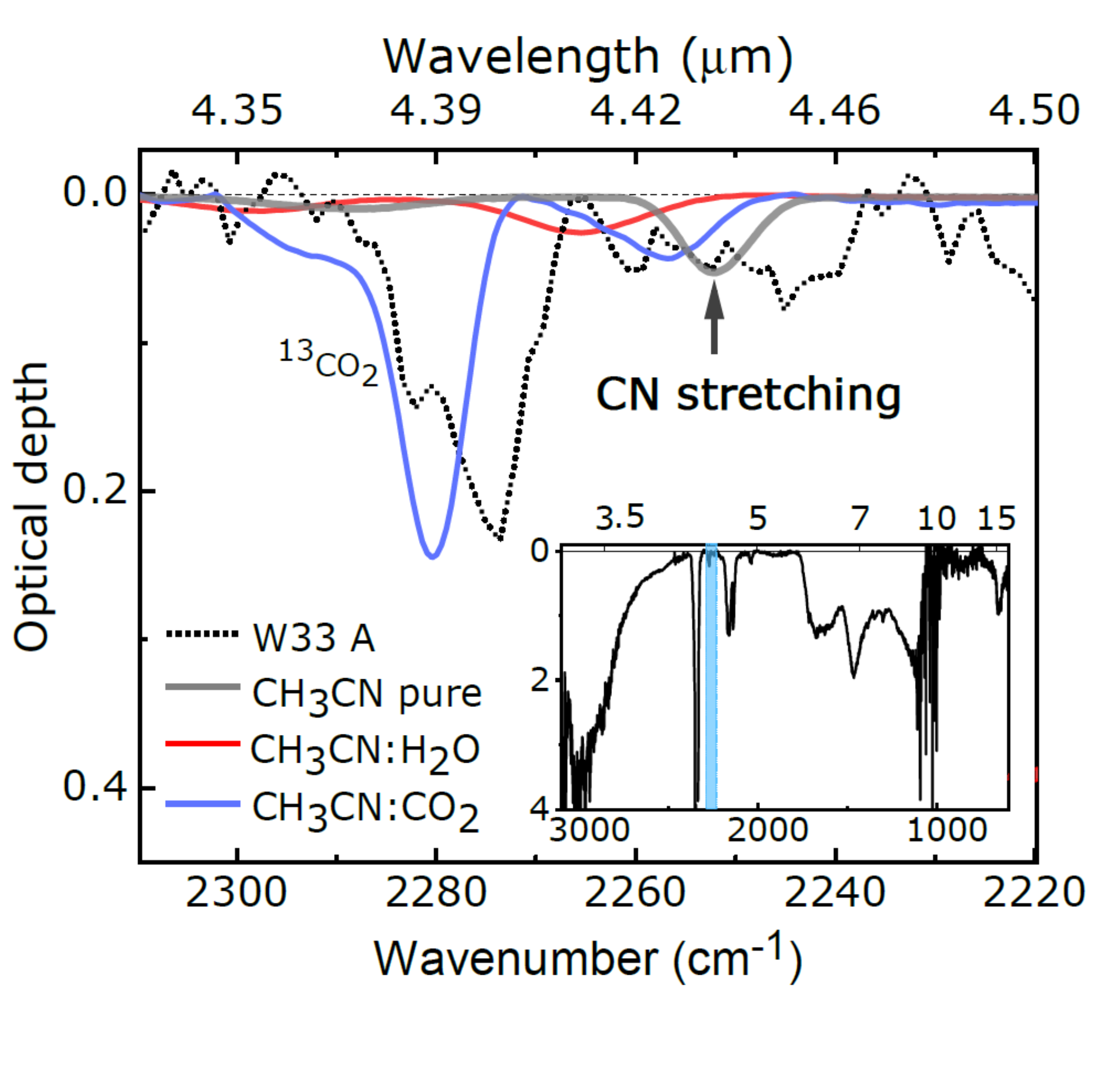}
\caption{ISO spectrum of W33A with subtraction of a local linear continuum between 4.48 - 4.35~$\mu$m (dotted line). The laboratory spectra of CH$_3$CN (gray), CH$_3$CN:H$_2$O (red) and CH$_3$CN:CO$_2$ (blue) at 15 K are also shown. The CN stretching of methyl cyanide ice in the laboratory spectrum, used for deriving upper limits for this species toward W33A, is indicated with an arrow. The inset in the
bottom right shows the complete spectrum of ices toward W33A. The region analyzed for deriving CH$_3$CN upper limits is indicated by the shadowed area.}
\label{upperlimit}
\end{figure}

\underline{Range between 6.0$-$10~$\mu$m}: Methyl cyanide has four bands in the region between 6.7 - 10  $\mu$m: at  6.905 $\mu$m (1448.3 cm$^{-1}$; assigned to a combination of vibrational modes), at 7.092 $\mu$m (1410 cm$^{-1}$; CH$_3$ antisymmetric deformation), and at 7.275 $\mu$m (1374.5 cm$^{-1}$; CH$_3$ symmetric deformation). The three features between 6.7$-$7.5~$\mu$m  are strong, but they overlap with the CH$_3$ deformation mode from other COMs \citep[e.g.,][]{Boogert2008}, and potential organic refractory residue \citep{Gibb2002, boogert2015observations, Rocha2021}. The CH$_3$ rock mode of methyl cyanide, around 9.600~$\mu$m (1041.6~cm$^{-1}$), can be a potential tracer of CH$_3$CN. Although methanol also shows a vibrational mode near 9.60~$\mu$m, the methyl cyanide's CH$_3$ rock mode is a relatively strong band and peaks at lower wavelengths.

\underline{Range above 10~$\mu$m:}
Above 10 $\mu$m, methyl cyanide has one band around 10.87~$\mu$m (919.6 cm$^{-1}$) assigned to the C$-$C stretching vibration. This feature has some disadvantages for being used as a methyl cyanide tracer: it is not readily observed when low quantities of methyl cyanide ($\sim$ 10$^{15}$ molecules cm$^{-2}$) are mixed with H$_2$O. Since this feature peaks in a spectral region dominated by strong absorption of H$_2$O and silicates, this band has limited potential as methyl cyanide tracer.

\begin{figure*}[ht]
\includegraphics[width=\hsize]{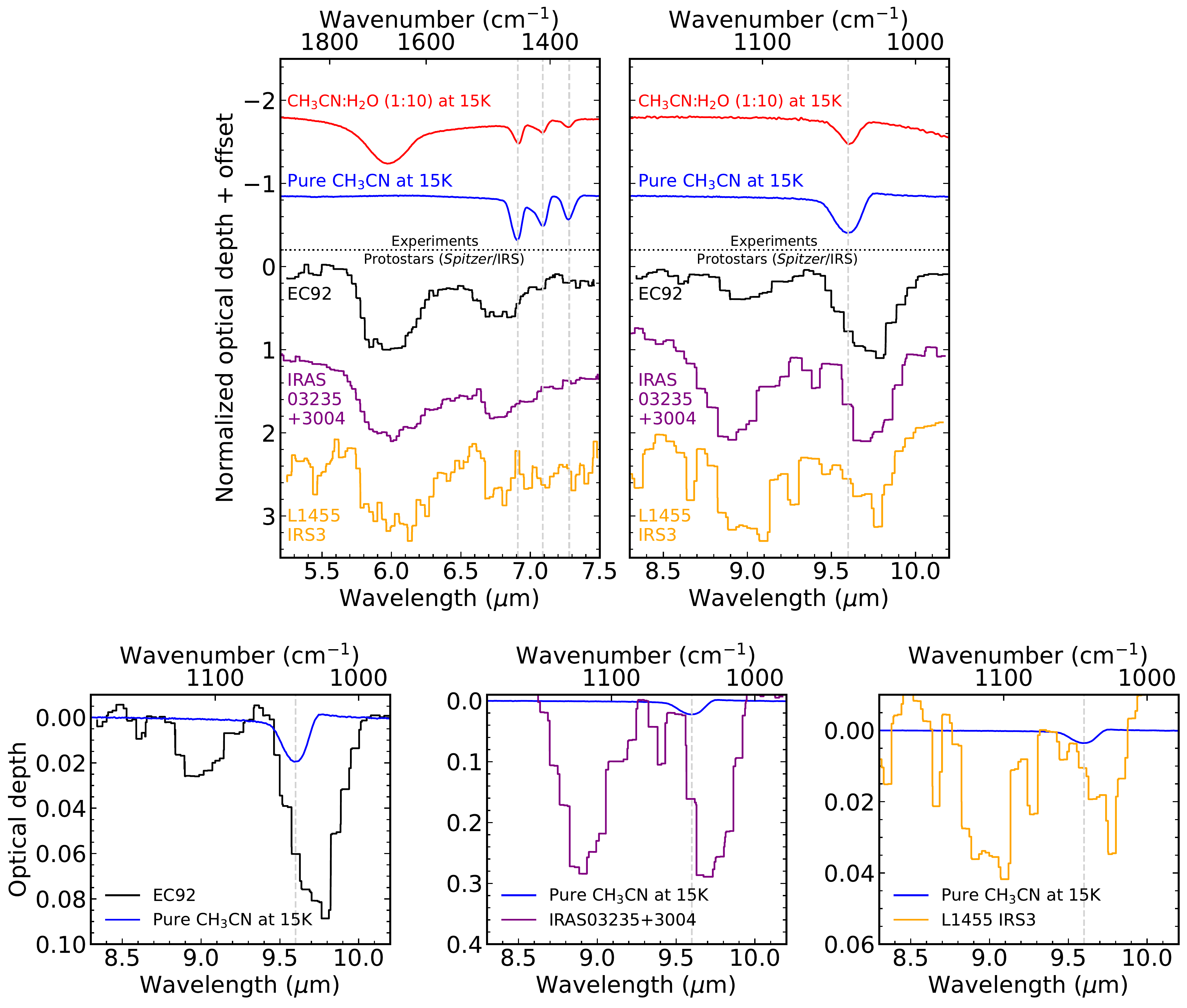}
\caption{Comparison between the normalized {\it Spitzer}/IRS spectrum of three YSOs (EC92, IRAS03235+3004, and L1455 IRS3) and laboratory spectra of CH$_3$CN ice. Top panels: the CH$_3$CN  and CH$_3$CN:H$_2$O ice spectra at 15~K are plotted with the YSO spectra (the continuum, and contributions from H$_2$O, and silicate features are subtracted). The 6.90, 7.09, 7.27, 9.60 $\mu$m features of CH$_3$CN are marked with a vertical dashed line. Bottom panels: {\it Spitzer}/IRS spectrum of EC92, IRAS03235+3004, and L1455 IRS3(from left to right) in the 8.5 - 10 $\mu$m region. In addition, the laboratory spectrum of pure CH$_3$CN ice at 15 K (blue) is shown. In the bottom panels, the spectrum of CH$_3$CN is scaled to the observations using the 9.60 $\mu$m band of methyl cyanide.}
\label{observations1500}
\end{figure*}

In this paper, we have searched for signatures of frozen CH$_3$CN toward four protostars. The first source is W33A, a well-known high-mass protostar \citep{gibb2000inventory, Gibb2004} that was observed with ISO, and thus has broad spectral coverage. A direct comparison of the laboratory spectra and the ISO spectrum shows that there is no convincing signal that would act as a basis for a (preliminary) identification of CH$_3$CN. However,  it is possible to derive an upper limit for the methyl cyanide column density using the data collected in this work. Figure \ref{upperlimit} shows the spectrum of W33A around the 4.4 $\mu$m region plotted with laboratory spectra of pure CH$_3$CN, CH$_3$CN:H$_2$O (1:10), and  CH$_3$CN:CO$_2$(1:10) ices, all at 15 K. To calculate the maximum content of solid CH$_3$CN toward W33A, the optical depth of the CN stretching feature (4.44~$\mu$m) in the laboratory spectra is visually scaled to the observational signal in this region. Using the integrated optical depth of the scaled laboratory feature and the CN stretching strength for pure and mixed ices at 15 K derived in this work, we obtain an upper limit for the CH$_3$CN column density of 2.4 $\times$ 10$^{17}$ molecules cm$^{-2}$, that corresponds to 1.9\% relative to the solid H$_2$O toward this object ($N_{\rm{H_2O}}$ $\sim$ 12.57 $\times$ 10$^{18}$; \citealt{Boogert2008}). By scaling the spectrum of the CH$_3$CN:H$_2$O(1:10) and the CH$_3$CN:CO$_2$(1:10) to the observations, the CH$_3$CN upper limit is around 1 $\times$ 10$^{17}$ molecules cm$^{-2}$, which corresponds to 0.8\% relative to H$_2$O ice.

\begin{figure}[h]
\centering
\includegraphics[width=1\hsize]{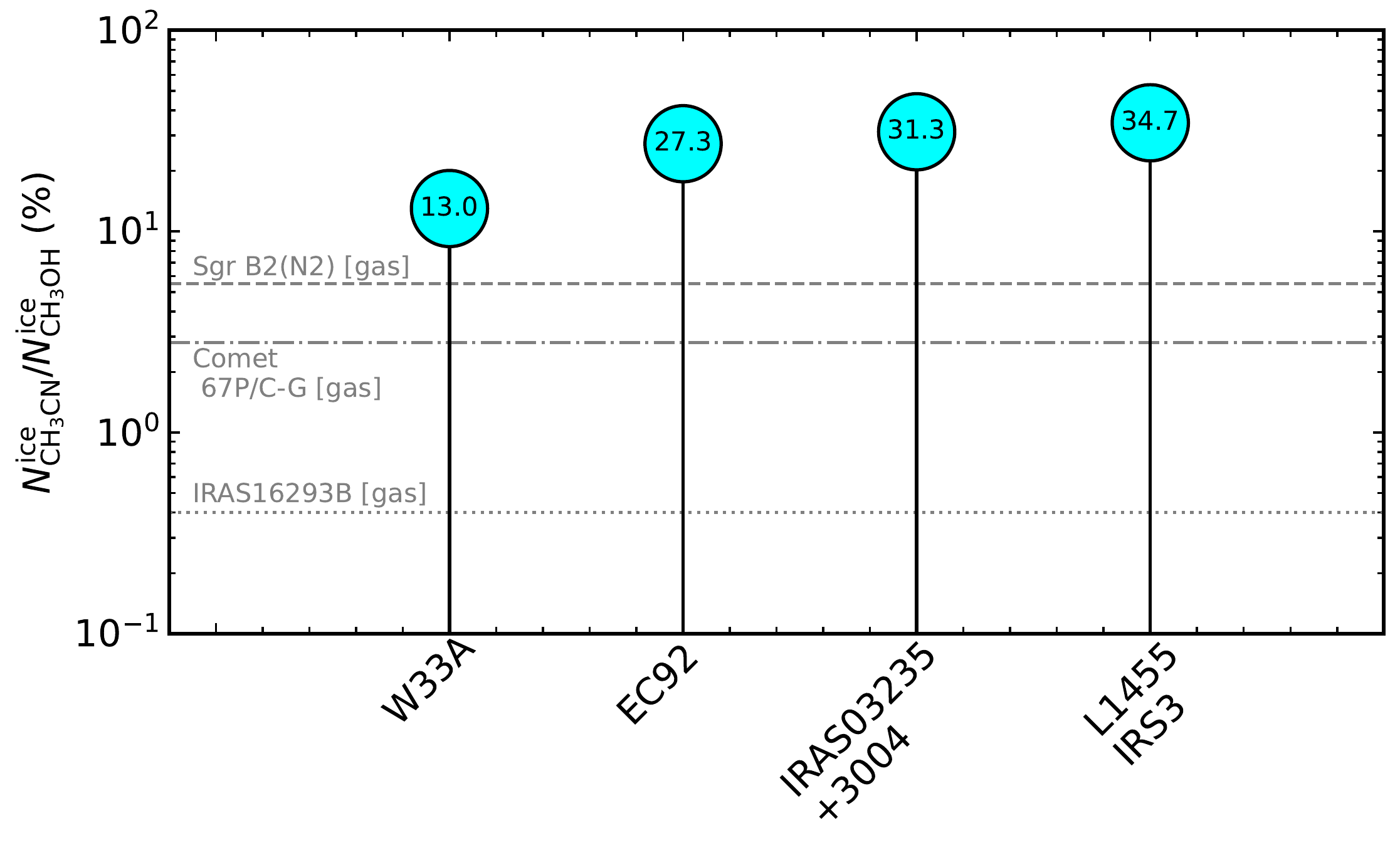}
\caption{Upper limit abundances in the solid phase of CH$_3$CN with respect to CH$_3$OH estimated for ices toward  W33A, EC92, IRAS03235+3004, and L1455 IRS3 derived in this work. The horizontal dashed lines show the gas-phase values for the hot molecular core Sgr~B2(N2), the Comet 67P/C-G coma, and the binary protostar IRAS 16293B (from \citealt{Jorgensen2020}).}
\label{ch3cn_ch3oh_abundances}
\end{figure}

In the top panel of Figure \ref{observations1500}, we show a comparison between the laboratory spectra of CH$_3$CN ice with the spectra of three low-mass protostars, namely, EC92, IRAS03325, and L1455 IRS3, acquired with the {\it Spitzer} telescope \citep{Boogert2008,bottinelli2010c2d}. The spectral coverage of these observations is between 5.3 and 38~$\mu$m, and therefore the region in which the CN stretching absorbs, around 4.44~$\mu$m, cannot be analyzed. As previously discussed, the three features between 6.7$-$7.5~$\mu$m (top left) cannot be used to constrain the CH$_3$CN abundances in the protostar spectra due to overlap with other ice features. For this reason, we use the CH$_3$ rock band of methyl cyanide, around 9.6 $\mu$m (top right), to estimate the upper limits for this molecule in the low-mass protostars observations. We stress that NH$_3$ and CH$_3$OH also absorb around this region \citep{bottinelli2010c2d}. However, the CH$_3$ rock band of methyl cyanide peaks at shorter wavelengths than the methanol band and it is, in this case, the best feature to estimate upper limits. The bottom panels of Figure~\ref{observations1500} show the comparison of the laboratory data presented here and the astronomical observations.  The absorption profile of the CH$_3$ rock of the pure methyl cyanide ice at 15~K is plotted superimposed to the spectra of the three YSOs. In each panel, the laboratory data are scaled to the observations to estimate the upper limits for methyl cyanide column density. The low resolution of the observations (R $\sim$ 100) and partial overlap with the methanol absorption feature prohibits distinguishing the different CH$_3$CN band profiles characterized in this work. For this reason, only the spectrum of pure methyl cyanide at 15 K is employed. The estimated upper-limits for methyl cyanide column density in these objects are 5.2 $\times$ 10$^{16}$, 1.9 $\times$ 10$^{17}$, and 8.5 $\times$ 10$^{17}$ molecules cm$^{-2}$ for  EC92,  IRAS03235+3004, and L1455 IRS3, respectively. Relative to H$_2$O ice, these upper limits correspond to abundances of  3.1, 1.3, and 4.1\% for EC92,  IRAS03235+3004, and L1455 IRS3, respectively (considering the H$_2$O column densities from \citealt{Boogert2008}).

Figure \ref{ch3cn_ch3oh_abundances} graphically displays the upper limit abundances of frozen methyl cyanide relative to solid-phase methanol derived for W33A, EC92, IRAS0325, and L1455 IRS3. The methanol ice column densities are taken from \citet{Boogert2008}. The abundances range from 13\% to 34.7\% and are higher than the CH$_3$CN/CH$_3$OH ratios observed in the gas-phase toward the hot molecular core Sgr~B2(N2), the Class 0 protostellar binary IRAS~16293B, and the coma of the Comet 67P/C-G  \citep[see Table~2 in][]{Jorgensen2020}. Despite the large uncertainties in the abundance calculation in the solid phase, these numbers suggest that the methyl cyanide content relative to methanol in interstellar ices can be larger than in the gas phase. An explanation would possibly pass through the thermal and nonthermal mechanisms releasing these molecules to the gas phase, their fragmentation upon desorption, and their full gas-phase reaction network. However, more sensitive ice observations are needed to confirm this trend.

Future observations combining the NIRspec \citep[Near-InfraRed Spectrograph;][]{Bagnasco2007} and MIRI \citep[Mid-InfraRed Instrument;][]{Wells2015} instruments on-board of JWST will cover the spectral regions where CH$_3$CN can be observed, including the band at 4.440 $\mu$m. A multiwavelength detection of methyl cyanide will help to understand the abundance of nitrogen-containing COMs in ices, which is currently less constrained than the O-containing COMs budget \citep{nazari2021complex}. The interpretation of future observations using laboratory spectra of CH$_3$CN in different ice matrices and temperatures will enable not only the identification of this species but also hint at the environment in which methyl cyanide is formed. The CH$_3$CN nondetection will also open questions regarding the origin of its high gas-phase abundance in hot cores and protostars.

\section{Conclusions}

This work reports a series of mid-infrared spectra measurements of methyl cyanide ice in its pure form and mixed with relevant interstellar ice molecules. The peak position and full width at half maximum of six methyl cyanide absorption bands are characterized in different ice mixtures and for temperatures ranging from 15 K up to 150 K. The refractive index and infrared band strengths of pure methyl cyanide, and relative band strengths for CH$_3$CN modes in ice mixtures, all measured at 15 K, are presented. The main conclusions of this work are the following:

\begin{enumerate}

\item For upcoming JWST observations, the best chance to identify frozen CH$_3$CN is through its CN stretching mode, which peaks at 2252.2 cm$^{-1}$ (4.440 $\mu$m) in the pure ice. This band together with the 1041.6 cm$^{-1}$ (9.600 $\mu$m; CH$_3$ rock) feature offers the most suitable combination of mid-IR transitions for identifying methyl cyanide in astronomical ice data. Due to the overlap with other interstellar ice components, the methyl cyanide bands at 2940.9 cm$^{-1}$ (3.400 $\mu$m; CH$_3$ symmetric stretching), 1448.3 cm$^{-1}$ (6.905 $\mu$m; CH$_3$CN combination of modes), 1410 cm$^{-1}$ (7.092 $\mu$m; CH$_3$ antisymmetric deformation), 1374.5 cm$^{-1}$ (7.275 $\mu$m ; CH$_3$ symmetric deformation), and (10.8 $\mu$m; C-C stretching) are less suited. However, the presence of bands in these regions should be checked as an additional tool to identify frozen CH$_3$CN.

\item Most of the analyzed infrared bands of CH$_3$CN in the ice mixtures shift by less than 5 cm$^{-1}$ with respect to the position in the pure ice. The most striking variations are observed for the CN stretching mode (2252 cm$^{-1}$) in H$_2$O-containing ices, in which the absorption peak appears around 2265 cm$^{-1}$. The changes in the FWHM when methyl cyanide is mixed with other molecules are more pronounced, and these are described throughout the work. Above 120 K, the peak position and FWHM of methyl cyanide bands in the mixtures are similar to the pure ice at the same temperature. This indicates that segregation processes may take place in the ice mixtures above 120 K.

\item The comparison between the recorded CH$_3$CN ice spectra with ISO observations of W33A in the 4 - 5 $\mu$m region yields an upper limit for its column density of  2.4 $\times$ 10$^{17}$ molecules cm$^{-2}$. Compared to H$_2$O and CH$_3$OH ice, this value corresponds to relative abundances of $\leq$ 1.9 and 13.0 percent, respectively.

\item The analysis of Spitzer/IRS spectra of ices toward the protostars EC92, IRAS 03235, and L1455 IRS3 in the 1800 - 1000 cm$^{-1}$ (5.5 - 10  $\mu$m)  region do not allow for the identification of CH$_3$CN features. Using the 1041.6 cm$^{-1}$ band of CH$_3$CN, an upper limit for the column density of this species is estimated as  5.2 $\times$ 10$^{16}$, 1.9 $\times$ 10$^{17}$ , and 3.8 $\times$ 10$^{16}$ molecules cm$^{-2}$ for EC92, IRAS 03235, and L1455 IRS3, respectively. With respect to solid H$_2$O, these values translate to relative abundances of 3.1, 1.3, and 4.1 percent, and with respect to methanol, to 27.3, 31.3, and 34.7 percent.
 
\end{enumerate}

The upcoming JWST observations of interstellar ices will reveal new features of solid-state molecules. Within several observing programs, including the JWST early release program Ice Age, several objects at the early stages of stellar evolution are targeted to search for COMs. The work presented here provides essential laboratory data to support the identification and quantification of CH$_3$CN in the observations. The identification of frozen methyl cyanide in the astronomical data will help understand the inventory and distribution of nitrogen-bearing molecules in star-forming regions.

\begin{acknowledgements}
  This work has been made possible through financial support by NOVA, the Netherlands  Research School for Astronomy, and NWO through its Dutch Astrochemistry Program (DAN II). We thank the anonymous referee for their very constructive suggestions and Dr. Thanja Lamberts for the interesting conversations about methyl cyanide chemistry. WRMR thanks the Leiden Observatory for its financial support. This work has been performed with support from the ICE AGE team.

\end{acknowledgements}


\bibliographystyle{aa}
\bibliography{biblio}

\onecolumn
\begin{appendix}

\section{Infrared profile of CH$_3$CN absorption bands in pure and mixed ices}
\import{./}{AppendixA-spectra.tex}

\clearpage

\section{Peak position and FWHM of CH$_3$CN bands in pure and mixed ices}
\import{./}{AppendixB-tables.tex}

\clearpage

\section{Integrated absorbance of CH$_3$CN bands in pure and mixed ices}
\import{./}{AppendixC-areas.tex}

\end{appendix}

\end{document}

%% file: AppendixA-spectra.tex
This section presents the infrared spectra of methyl cyanide-containing ices in pure and mixed forms. Each figure displays the profile of one band in different ice mixtures (shown in the different panels) and at different temperatures. The peak position and FWHM for selected temperatures are shown in the bottom left panels of each figure. The relative band strengths of the analyzed absorption band in the different ices at 15 K are graphically presented in the bottom right panels.

\begin{figure*}[ht]
\includegraphics[width=1.0\linewidth]{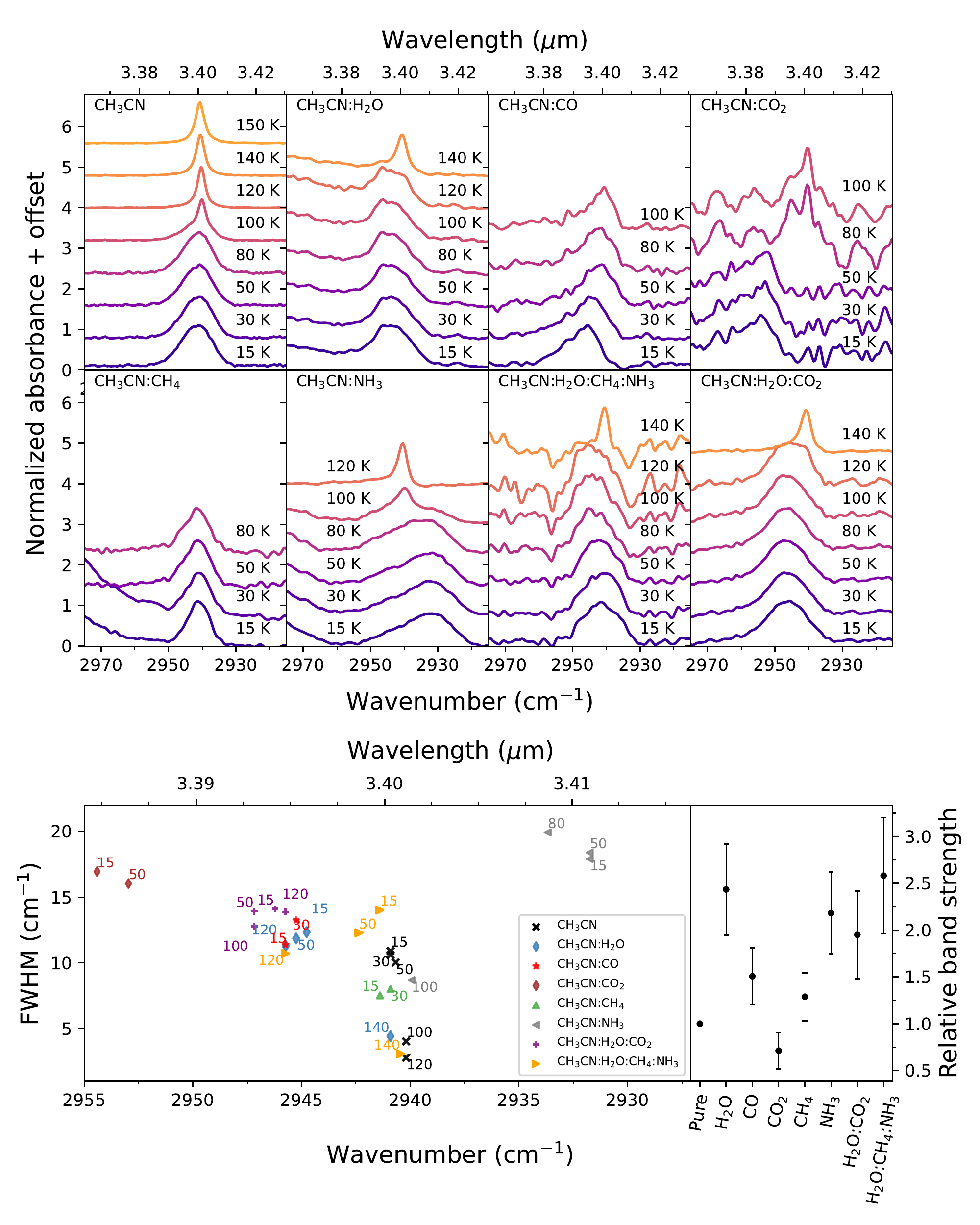}
\caption{Upper panel: Absorption profile of the C-H symmetric stretch mode of methyl cyanide, around 2940.9 cm$^{-1}$ (3.400~$\mu$m), in pure and mixed ices. The ice spectra at different temperatures are indicated by different colors and labels. Bottom left: Peak position and FWHM of the C-H symmetric stretching mode of CH$_3$CN in different ice mixtures at selected temperatures. Bottom right: relative band strengths of the C-H symmetric stretch band in different ice mixtures at 15 K.}
\label{fig:2900}
\end{figure*}
\clearpage

\begin{figure*}[ht]
\includegraphics[width=1.0\linewidth]{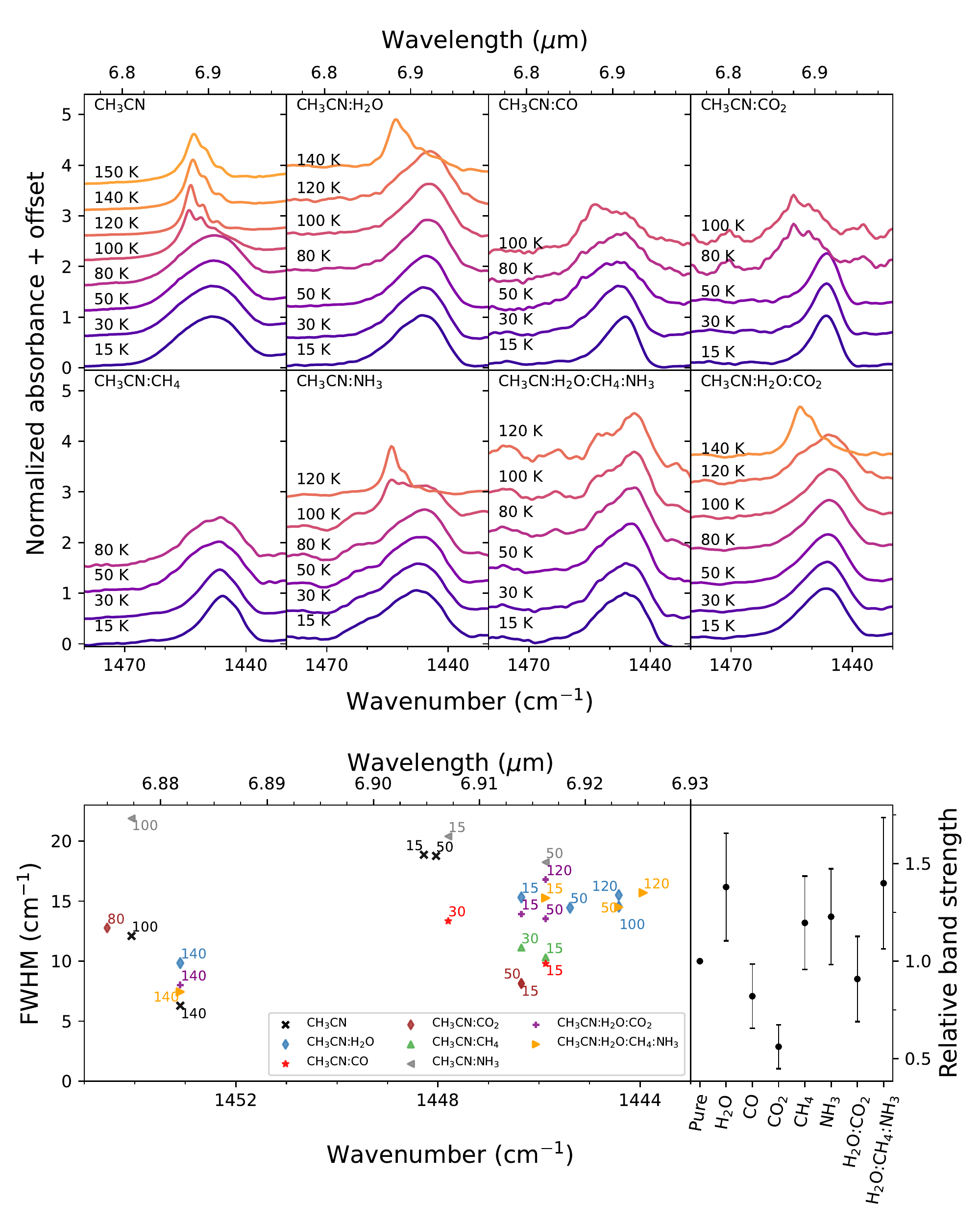}
\caption{Upper panel: Absorption profile of the combination mode of methyl cyanide, around 1448.3 cm$^{-1}$ (6.905 $\mu$m), in pure and mixed ices. The ice spectra at different temperatures are indicated by different colors and labels. Bottom left: Peak position and FWHM of the combination mode of CH$_3$CN in different ice mixtures at selected temperatures. Bottom right: relative band strengths  of the combination band in different ice mixtures at 15 K.}
\label{fig:1450}
\end{figure*}
\clearpage

\begin{figure*}[ht]
\includegraphics[width=1.0\linewidth]{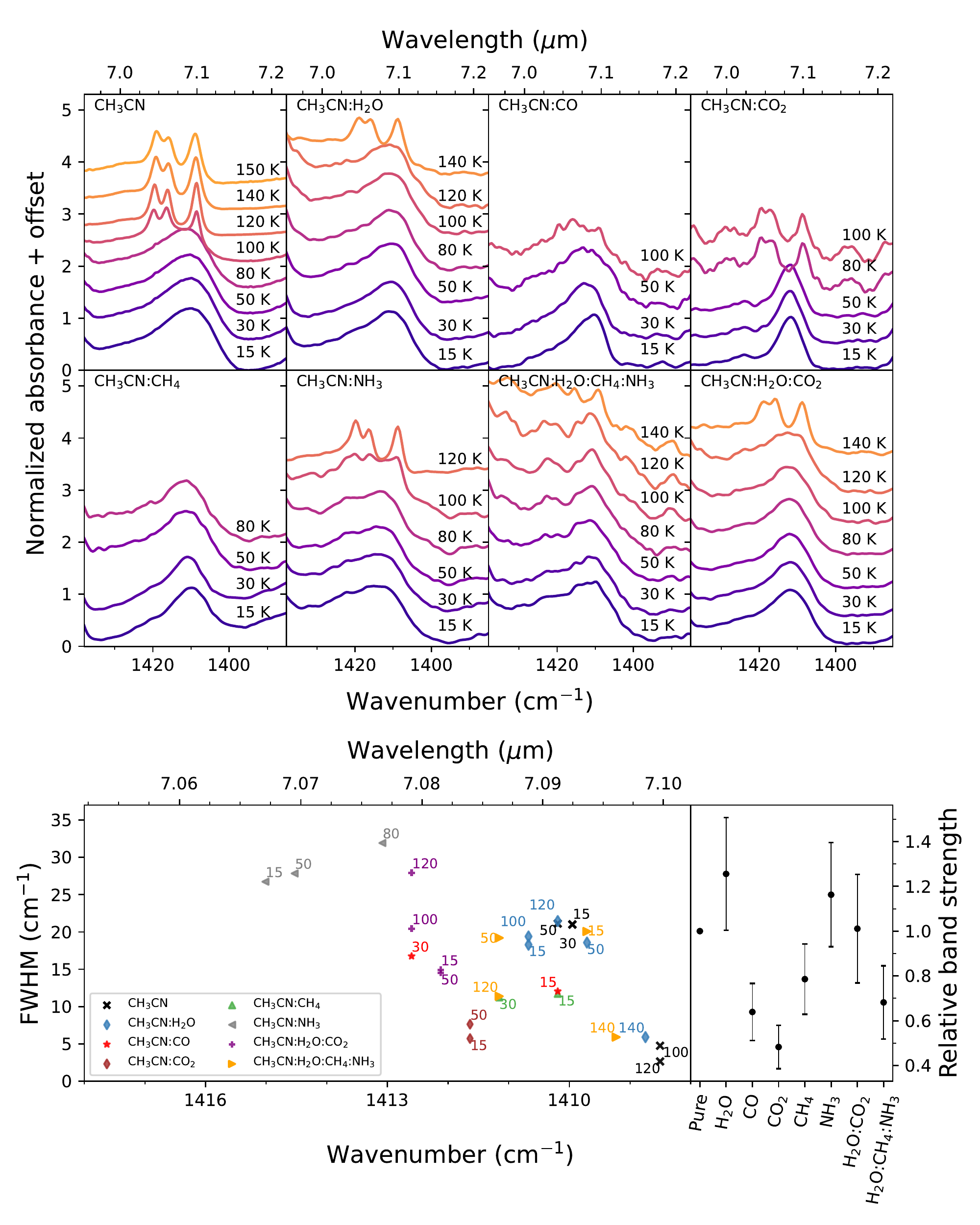}
\caption{Upper panel: Absorption profile of the CH$_3$ antisymmetric deformation mode of methyl cyanide, around 1410 cm$^{-1}$ (7.092 $\mu$m), in pure and mixed ices. The ice spectra at different temperatures are indicated by different colors and labels. Bottom left: Peak position and FWHM of the CH$_3$ antisymmetric deformation mode of CH$_3$CN in different ice mixtures at selected temperatures. Bottom right: relative band strengths  of the CH$_3$ antisymmetric deformation band in different ice mixtures at 15 K.}
\label{fig:1400}
\end{figure*}
\clearpage

\begin{figure*}[ht]
\includegraphics[width=1.0\linewidth]{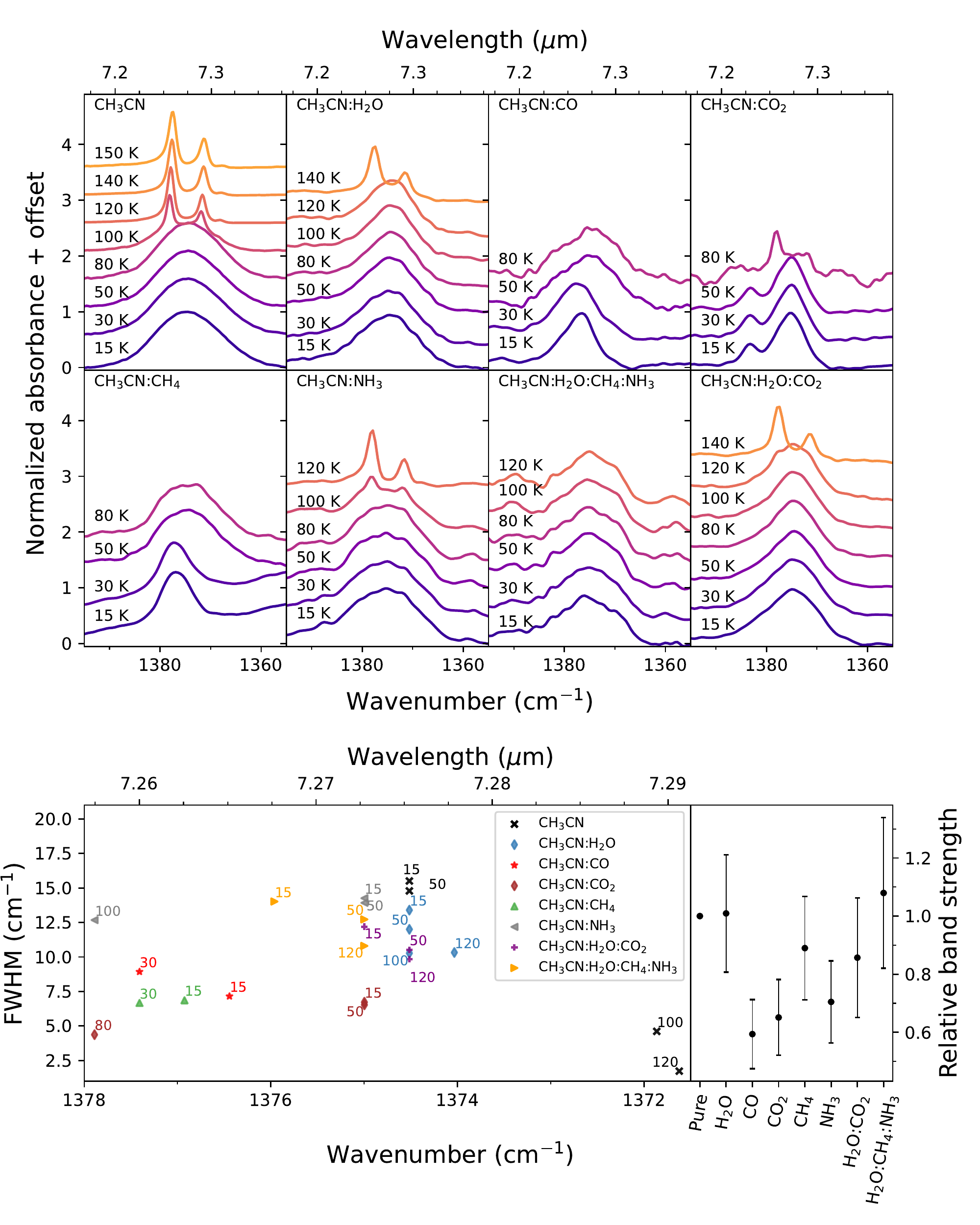}
\caption{Upper panel: Absorption profile of the CH$_3$ symmetric deformation mode of methyl cyanide, around 1374.5 cm$^{-1}$ (7.275 $\mu$m), in pure and mixed ices. The ice spectra at different temperatures are indicated by different colors and labels. Bottom left: Peak position and FWHM of the CH$_3$ symmetric deformation mode of CH$_3$CN in different ice mixtures at selected temperatures. Bottom right: relative band strengths  of the CH$_3$ symmetric deformation band in different ice mixtures at 15 K.}
\label{fig:1350}
\end{figure*}
\clearpage

\begin{figure*}[ht]
\includegraphics[width=1.0\linewidth]{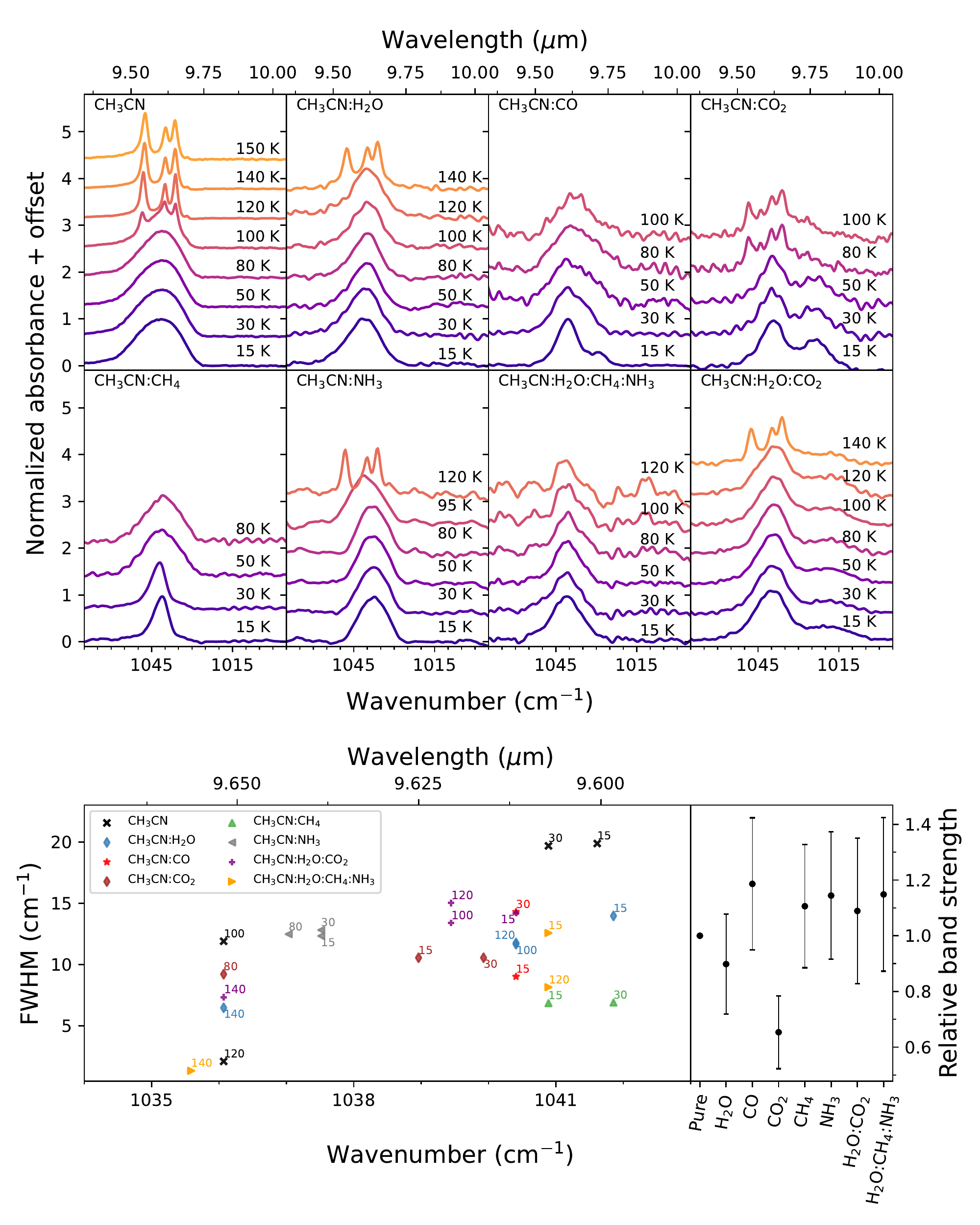}
\caption{Upper panel: Absorption profile of the CH$_3$ rock mode of methyl cyanide, around 1041.6 cm$^{-1}$ (9.600$\mu$m), in pure and mixed ices. The ice spectra at different temperatures are indicated by different colors and labels. Bottom left: Peak position and FWHM of the CH$_3$ rock mode of CH$_3$CN in different ice mixtures at selected temperatures. Bottom right: relative band strengths  of the CH$_3$ rock band in different ice mixtures at 15 K.}
\label{fig:1040}
\end{figure*}
\clearpage

%% file: AppendixB-tables.tex
In this section, the peak position and FWHM of the methyl cyanide features are presented. Each table lists the peak position and FWHM values for one band in the pure ice and mixed with H$_2$O, CO, CO$_2$, CH$_4$, NH$_3$, H$_2$O:CO$_2$ and H$_2$O:CH$_4$:NH$_3$.  When the analyzed absorption band splits into multiple components at higher temperatures, the peak position of all the components and the FWHM of the strongest component are given. The uncertainty in the peak position is 1 cm$^{-1}$ and the uncertainty in the FWHM amounts to 1.5 cm$^{-1}$. The values marked with an asterisk (*) denote peaks for which the FHWM is a combination of more than one peak.

\begin{table*}[h]
\centering
\setlength{\tabcolsep}{3.5pt} 
\renewcommand{\arraystretch}{1.1} 
\fontsize{9}{10}\selectfont 
\caption[]{Peak position and FWHM of methyl cyanide CH$_3$ symmetric stretching mode (2940.9 cm$^{-1}$/3.400 $\mu$m) in pure and mixed ices at temperatures ranging from 15  - 150 K.}       
\begin{threeparttable}[b]
\begin{tabular}{l|l|cc|cc|cc|cc|cc|cc}     
\hline           
\multirow{2}{*}{T} & \multirow{2}{*}{Matrix}&  \multicolumn{4}{c|}{1:5}& \multicolumn{4}{c|}{1:10}& \multicolumn{4}{c}{1:20}  \\

\cline{3-14}
 &   & \multirow{2}{*}{Peak}& \multirow{2}{*}{FWHM} & \multirow{2}{*}{Peak}& \multirow{2}{*}{FWHM} &\multirow{2}{*}{Peak}& \multirow{2}{*}{FWHM} &\multirow{2}{*}{Peak}& \multirow{2}{*}{FWHM}&\multirow{2}{*}{Peak}& \multirow{2}{*}{FWHM}&\multirow{2}{*}{Peak}& \multirow{2}{*}{FWHM} \\
&   &   &   & &   & &   & &&&   & & \\ 
&   & (cm$^{-1}$) &(cm$^{-1}$)  & ($\mu$m)& ($\mu$m) &(cm$^{-1}$) &(cm$^{-1}$)  & ($\mu$m)& ($\mu$m)&(cm$^{-1}$) &(cm$^{-1}$)  & ($\mu$m)& ($\mu$m)\\ 
 
\hline    
\multirow{5}{*}{15} &Pure ice& 2940.9& 10.9& 3.400& 0.013 & - &-  &  -& - & - & - & - &- \\
&H$_2$O & 2942.8& 12.5& 3.398& 0.014  & 2944.8& 12.2& 3.396& 0.014 & 2944.8& 11.8& 3.396& 0.014 \\
&CO& 2944.3& 7.4& 3.396& 0.009  & 2945.7& 11.7& 3.395& 0.013 & -  & -  & -  & -\\

&CO$_2$ & 2952.0& 16.6& 3.388& 0.019& 2954.4& 17.9& 3.385& 0.020& -  & -  & -  & -\\
&CH$_4$ & 2940.4& 8.7& 3.401& 0.010   & 2941.4& 7.5& 3.400& 0.009 & 2941.4& 8.6& 3.400& 0.010 \\

&NH$_3$ & 2934.2& 18.0& 3.408& 0.021& 2931.8& 18.0& 3.411& 0.021  & 2934.2& 15.2& 3.408& 0.018 \\
&H$_2$O:CO$_2$ & -  & -  & -  & -  & 2945.7& 13.9& 3.395& 0.016 & -  & -  & -  & -\\

& 4 comp& -  & -  & -  & -  & -  & -  & -  & -  &2941.4& 14.1& 3.340& 0.016  \\\hline    
\multirow{5}{*}{30} &Pure ice& 2940.9& 10.6& 3.400& 0.012  &-  &  -& - & - & - & - &- \\
&H$_2$O & 2942.4& 12.5& 3.399& 0.014 & 2943.8& 11.8& 3.397& 0.014& 2944.8& 12.1& 3.396& 0.014 \\
&CO& 2943.8& 7.9& 3.397& 0.009 & 2945.3& 13.3& 3.395& 0.015  & -  & -  & -  & -\\

&CO$_2$ & 2951.0& 13.9& 3.389& 0.016 & 2953.4& 17.3& 3.386& 0.020 & -  & -  & -  & -\\
&CH$_4$ & 2940.4& 8.3& 3.401& 0.010& 2940.9& 7.8& 3.400& 0.009 & 2940.4& 11.2& 3.401& 0.013 \\

&NH$_3$ & 2933.7& 16.8& 3.409& 0.020   & 2932.7& 17.2& 3.410& 0.020 & 2934.6& 15.3& 3.408& 0.018 \\
&H$_2$O:CO$_2$ & -  & -  & -  & -  & 2947.2& 13.9& 3.393& 0.016  & -  & -  & -  & -\\

& 4 comp& -  & -  & -  & -  & -  & -  & -  & -  &2940.4& 13.2& 3.401& 0.015  \\\hline   
\multirow{5}{*}{50} &Pure ice& 2940.7& 10.0& 3.401& 0.012  &-  &  -& - & - & - & - &- \\

&H$_2$O & 2942.8& 12.4& 3.398& 0.014 & 2945.3& 11.8& 3.395& 0.014 & 2944.3& 12.6& 3.396& 0.015 \\

&CO$_2$ & 2950.6& 13.9& 3.389& 0.016   &  2953.0& 16.1& 3.386& 0.018  & -  & -  & -  & -\\
&CH$_4$ & 2941.4& 9.5& 3.400& 0.011 & 2941.4& 8.3& 3.400& 0.009 & 2939.9& 11.2& 3.401& 0.013 \\

&NH$_3$& 2933.2& 17.3& 3.409& 0.020 &2931.8& 18.5& 3.411& 0.022  & 2936.1& 15.2& 3.406& 0.018 \\
&H$_2$O:CO$_2$ & -  & -  & -  & -  & 2947.7& 13.9& 3.392& 0.016  & -  & -  & -  & -\\

& 4 comp& -  & -  & -  & -  & -  & -  & -  & -  & 2942.4& 12.6& 3.399& 0.014 \\ \hline
\multirow{6}{*}{80} &Pure ice& 2940.9& 9.6& 3.400& 0.011 - &-  &  -& - & - & - & - &- \\

&H$_2$O & 2942.8& 12.0& 3.398& 0.014 & 2945.7& 11.2& 3.395& 0.013  & 2945.3& 10.9& 3.395& 0.013 \\
&CO$_2$& 2950.1& 15.2& 3.390& 0.017 & 2942.4& 11.1& 3.399& 0.013  & -  & -  & -  & -\\


&NH$_3$& 2934.6& 16.9& 3.408& 0.020  & 2933.7& 20.1& 3.409& 0.023 & 2936.6& 14.2& 3.405& 0.016 \\
&H$_2$O:CO$_2$ & -  & -  & -  & -  & 2945.7& 13.3& 3.395& 0.015  & -  & -  & -  & -\\

& 4 comp& -  & -  & -  & -  & -  & -  & -  & -  & 2945.3& 11.3& 3.395& 0.013 \\\hline

\multirow{5}{*}{100} &Pure ice& 2940.2& 4.0& 3.401& 0.005 & - &-  &  -& - & - & - & - &- \\

&H$_2$O & 2942.8& 11.8& 3.398& 0.014  & 2945.7& 11.2& 3.395& 0.013 & 2945.3& 10.4& 3.395& 0.012 \\

&CO$_2$ & 2940.9& 9.6& 3.400& 0.011   & -  & -  & -  & -  & -  & -  & -  & -\\

&NH$_3$ & 2939.9& 9.4& 3.401& 0.011 & 2939.9& 8.8& 3.401& 0.010 & 2938.0& 11.9& 3.404& 0.014 \\
&H$_2$O:CO$_2$ & -  & -  & -  & -  & 2947.7& 12.8& 3.392& 0.015  & -  & -  & -  & -\\

& 4 comp& -  & -  & -  & -  & -  & -  & -  & -  & 2944.3& 10.6& 3.396& 0.012 \\\hline
\multirow{8}{*}{120} &Pure ice& 2940.2& 2.8& 3.401& 0.003 & - &-  &  -& - & - & - & - &- \\

&H$_2$O & 2941.9& 11.1& 3.399& 0.013  & 2945.7& 11.2& 3.395& 0.013 & 2945.3& 9.4& 3.395& 0.011 \\

&NH$_3$ & 2940.4& 3.4& 3.401& 0.004  & 2940.4& 3.3& 3.401& 0.004& 2938.0& 11.9& 3.404& 0.014 \\

&H$_2$O:CO$_2$ & -  & -  & -  & -  & 2945.7& 14.0& 3.395& 0.016& -  & -  & -  & -\\

& 4 comp& -  & -  & -  & -  & -  & -  & -  & -  & 2945.7& 11.0& 3.395& 0.013 \\\hline
\multirow{8}{*}{140} &Pure ice& 2940.4& 3.0& 3.401& 0.003 & - &-  &  -& - & - & - & - &- \\

&H$_2$O & 2940.9& 4.6& 3.400& 0.005   & 2940.9& 4.4& 3.400& 0.005  & 2940.9& 4.9& 3.400& 0.006 \\

&NH$_3$ & -  & -  & -  & -  & -  & -  & -  & -  & -  & -  & -  & -\\
&H$_2$O:CO$_2$ & -  & -  & -  & -  & 2940.9& 3.9& 3.400& 0.005 & -  & -  & -  & -\\

& 4 comp& -  & -  & -  & -  & -  & -  & -  & -  & -  & -  & -  & -\\
\hline
\multirow{1}{*}{150} &Pure ice& 2940.7& 3.1& 3.401& 0.004 & - &-  &  -& - & - & - & - &- \\

\hline    
\end{tabular}
\centering
  \end{threeparttable}
  \label{2940-mixtures}
\end{table*}

\clearpage

\begin{table*}[h]
\centering
\setlength{\tabcolsep}{3.5pt} 
\renewcommand{\arraystretch}{1.1} 

\fontsize{9}{10}\selectfont 
\caption[]{Peak position and FWHM of methyl cyanide CN stretching mode (2252.2 cm$^{-1}$/ 4.440 $\mu$m) in pure and mixed ices at temperatures ranging from 15  - 150 K.}       
\begin{threeparttable}[b]
\begin{tabular}{l|l|cc|cc|cc|cc|cc|cc}     
\hline           
\multirow{2}{*}{T} & \multirow{2}{*}{Matrix}&  \multicolumn{4}{c|}{1:5}& \multicolumn{4}{c|}{1:10}& \multicolumn{4}{c}{1:20}  \\

\cline{3-14}
 &   & \multirow{2}{*}{Peak}& \multirow{2}{*}{FWHM} & \multirow{2}{*}{Peak}& \multirow{2}{*}{FWHM} &\multirow{2}{*}{Peak}& \multirow{2}{*}{FWHM} &\multirow{2}{*}{Peak}& \multirow{2}{*}{FWHM}&\multirow{2}{*}{Peak}& \multirow{2}{*}{FWHM}&\multirow{2}{*}{Peak}& \multirow{2}{*}{FWHM} \\
&   &   &   & &   & &   & &&&   & & \\ 
&   & (cm$^{-1}$) &(cm$^{-1}$)  & ($\mu$m)& ($\mu$m) &(cm$^{-1}$) &(cm$^{-1}$)  & ($\mu$m)& ($\mu$m)&(cm$^{-1}$) &(cm$^{-1}$)  & ($\mu$m)& ($\mu$m)\\ 
 
\hline    
\multirow{5}{*}{15}&Pure ice & 2252.2& 8.0& 4.440& 0.016  & - &-  &  -& - & - & - & - &- \\

&H$_2$O& 2265.0& 13.2& 4.415& 0.026  & 2265.5& 13.2& 4.414& 0.026 & 2266.4& 14.4& 4.412& 0.028  \\

&CO& 2254.4& 4.6& 4.436& 0.009 & 2253.9& 4.8& 4.437& 0.009 & -  & -  & -  & -\\

&CO$_2$& 2255.8& 9.4& 4.433& 0.018 & 2256.8& 11.7& 4.431& 0.023  & -  & -  & -  & -\\

&CH$_4$& 2254.4& 5.4& 4.436& 0.011 & 2254.9& 5.2& 4.435& 0.010  & 2255.8& 4.8& 4.433& 0.009   \\

&NH$_3$& 2252.4& 8.3& 4.440& 0.016 & 2252.0& 8.3& 4.441& 0.016  & 2252.4& 8.8& 4.440& 0.017\\
&H$_2$O:CO$_2$ & - &-  &  -& -   & 2263.1& 15.0& 4.419& 0.029 & -  & -  & -  & -\\

& 4 comp& - &-  &  -& - & - & - & - &-  &2264.0& 13.6& 4.417& 0.026    \\\hline    

\multirow{5}{*}{30}&Pure ice & 2252.2& 7.8& 4.440& 0.015 & - &-  &  -& - & - & - & - &- \\
&H$_2$O& 2265.0& 13.0& 4.415& 0.025&  2265.9& 13.1& 4.413& 0.025  & 2265.9& 14.5& 4.413& 0.028   \\

&CO& 2253.9& 4.6& 4.437& 0.009 & 2253.9& 4.8& 4.437& 0.009  & -  & -  & -  & -\\

&CO$_2$& 2255.8& 9.2& 4.433& 0.018& 2256.8& 11.1& 4.431& 0.022 & -  & -  & -  & -\\

&CH$_4$& 2254.4& 5.2& 4.436& 0.010 & 2254.4& 4.8& 4.436& 0.009  & 2254.9& 4.8& 4.435& 0.009   \\

&NH$_3$& 2252.4& 8.3& 4.440& 0.016 & 2252.0& 8.2& 4.441& 0.016  & 2252.0& 8.7& 4.441& 0.017\\

&H$_2$O:CO$_2$  & - &-  &  -& -   & 2263.1& 14.9& 4.419& 0.029 & -  & -  & -  & -\\

& 4 comp& - &-  &  -& - & - & - & - &- & 2264.0& 13.8& 4.417& 0.027  \\\hline

\multirow{5}{*}{50}&Pure ice & 2252.0& 7.5& 4.441& 0.015  & - &-  &  -& - & - & - & - &-  \\
&H$_2$O& 2265.5& 12.4& 4.414& 0.024 & 2266.4& 12.4& 4.412& 0.024 & 2265.9& 13.7& 4.413& 0.027   \\

&CO$_2$& 2255.8& 8.7& 4.433& 0.017 & 2256.8& 11.3& 4.431& 0.022 & -  & -  & -  & -\\

&CH$_4$& 2252.9& 6.8& 4.439& 0.013 & 2252.4& 6.8& 4.440& 0.013 & 2252.9& 7.1& 4.439& 0.014   \\

&NH$_3$& 2252.4& 8.0& 4.440& 0.016 & 2252.0& 8.0& 4.441& 0.016 & 2252.0& 8.4& 4.441& 0.017 \\
&H$_2$O:CO$_2$  & - &-  &  -& -  & 2263.1& 14.5& 4.419& 0.028  & -  & -  & -  & -\\

& 4 comp & - &-  &  -& - & - & - & - &-  & 2264.0& 13.2& 4.417& 0.026   \\ \hline
\multirow{5}{*}{80}&Pure ice & 2252.0& 7.1& 4.441& 0.014  & - &-  &  -& - & - & - & - &-  \\
 &H$_2$O& 2265.9& 11.5& 4.413& 0.022  & 2266.4& 11.7& 4.412& 0.023 & 2266.9& 12.3& 4.411& 0.024 \\
 

&CO$_2$ & 2255.3& 8.0& 4.434& 0.016 & 2251.0& 3.3& 4.442& 0.006  & -  & -  & -  & -\\


&NH$_3$& 2252.4& 7.5& 4.440& 0.015& 2252.4& 7.5& 4.440& 0.015  & 2252.4& 7.9& 4.440& 0.016\\
&H$_2$O:CO$_2$  & - &-  &  -& -  & 2263.1& 14.5& 4.419& 0.028 & -  & -  & -  & -\\

& 4 comp   & - &-  &  -& - & - & - & - &- & 2266.4& 12.5& 4.412& 0.024    \\\hline

\multirow{5}{*}{100}&Pure ice & 2251.0& 2.7& 4.4425& 0.0053  & - &-  &  -& - & - & - & - &-  \\
&H$_2$O& 2265.5& 11.1& 4.414& 0.022& 2266.4& 11.1& 4.412& 0.022 & 2267.4& 12.0& 4.410& 0.023   \\

&CO$_2$& 2251.0& 5.7& 4.442& 0.011 & 2251.0& 3.2& 4.442& 0.006  & -  & -  & -  & -\\

&NH$_3$& 2251.5& 6.3& 4.441& 0.012 & 2251.0& 2.1& 4.442& 0.004  & -  & -  & -  & -\\
&H$_2$O:CO$_2$  & - &-  &  -& -   & 2265.0& 12.4& 4.415& 0.024  & -  & -  & -  & -\\

& 4 comp & - &-  &  -& - & - & - & - &- & 2266.4& 12.2& 4.412& 0.024    \\\hline

\multirow{3}{*}{120}&Pure ice & 2251.0& 1.3& 4.442& 0.003 & - &-  &  -& - & - & - & - &- \\
&H$_2$O& 2265.5& 11.1& 4.414& 0.022  & 2266.4& 11.2& 4.412& 0.022  & 2266.4& 11.0& 4.412& 0.021   \\
&NH$_3$& -  & & -  & & -  & -  & -  & -  & 2252.4& 8.4& 4.440& 0.017\\
&H$_2$O:CO$_2$  & - &-  &  -& -   & 2265.0& 12.3& 4.415& 0.024  & -  & -  & -  & -\\
& 4 comp& - &-  &  -& - & - & - & - &-  & 2266.4& 12.5& 4.412& 0.024   \\\hline

\multirow{3}{*}{140}&Pure ice & 2250.8& 1.3& 4.443& 0.003 & - &-  &  -& - & - & - & - &- \\
 &H$_2$O& 2251.0& 2.0& 4.442& 0.004 & 2251.0& 2.0& 4.442& 0.004  & 2251.0& 2.0& 4.442& 0.004   \\

&NH$_3$& 2251.0& 2.1& 4.442& 0.004& -  & -  & -  & -  & -  & -  & -  &   \\
&H$_2$O:CO$_2$  & - &-  &  -& -  & 2251.0& 2.1& 4.442& 0.004  & -  & -  & -  & -\\

& 4 comp & - &-  &  -& - & - & - & - &- & 2251.0& 2.1& 4.442& 0.004     \\
\hline    
\end{tabular}
\centering
  \end{threeparttable}
  \label{2252-mixtures}
\end{table*}

\begin{table*}[h]
\centering
\setlength{\tabcolsep}{3.5pt} 
\renewcommand{\arraystretch}{1.1} 

\fontsize{9}{10}\selectfont 
\caption[]{Peak position and FWHM of methyl cyanide combination mode (1448.3 cm$^{-1}$/ 6.905 $\mu$m) in pure and mixed ices at temperatures ranging from 15  - 150 K.}       
\begin{threeparttable}[b]
\begin{tabular}{l|l|cc|cc|cc|cc|cc|cc}     
\hline           
\multirow{2}{*}{T} & \multirow{2}{*}{Matrix}&  \multicolumn{4}{c|}{1:5}& \multicolumn{4}{c|}{1:10}& \multicolumn{4}{c}{1:20}  \\

\cline{3-14}
 &   & \multirow{2}{*}{Peak}& \multirow{2}{*}{FWHM} & \multirow{2}{*}{Peak}& \multirow{2}{*}{FWHM} &\multirow{2}{*}{Peak}& \multirow{2}{*}{FWHM} &\multirow{2}{*}{Peak}& \multirow{2}{*}{FWHM}&\multirow{2}{*}{Peak}& \multirow{2}{*}{FWHM}&\multirow{2}{*}{Peak}& \multirow{2}{*}{FWHM} \\
&   &   &   & &   & &   & &&&   & & \\ 
&   & (cm$^{-1}$) &(cm$^{-1}$)  & ($\mu$m)& ($\mu$m) &(cm$^{-1}$) &(cm$^{-1}$)  & ($\mu$m)& ($\mu$m)&(cm$^{-1}$) &(cm$^{-1}$)  & ($\mu$m)& ($\mu$m)\\ 
 
\hline    
\multirow{5}{*}{15} &Pure ice & 1448.3& 18.9& 6.9045& 0.090 & - &-  &  -& - & - & - & - &- \\

&H$_2$O & 1446.4& 15.9& 6.914& 0.076 & -  & -  & -  & -  & 1446.4& 15.0& 6.914& 0.072 \\
&CO& 1447.3& 10.7& 6.909& 0.051 & 1445.9& 9.8& 6.916& 0.047 & -  & -  & -  & -\\

&CO$_2$ & 1446.8& 9.8& 6.911& 0.047  &  1446.4& 8.1& 6.914& 0.038   & -  & -  & -  & -\\

&CH$_4$ & 1446.4& 11.5& 6.914& 0.055 & 1445.9& 10.3& 6.916& 0.049 & 1445.4& 9.6& 6.919& 0.046 \\

&NH$_3$ & 1446.4& 19.8& 6.914& 0.095 & 1447.8& 20.4& 6.907& 0.097  & 1446.8& 17.9& 6.912& 0.086 \\
&H$_2$O:CO$_2$ & -  & -  & -  & -  & 1446.4& 13.9& 6.914& 0.066 & -  & -  & -  & -\\

& 4 comp  & -  & -  & -  & -  & -  & -  & -  & -  & 1445.9& 15.3& 6.916& 0.073 \\\hline    
\multirow{5}{*}{30} &Pure ice & 1448.5& 18.8& 6.904& 0.090  & - &-  &  -& - & - & - & - &- \\

&H$_2$O & 1446.4& 15.6& 6.914& 0.075 & -  & -  & -  & -  & 1444.4& 14.6& 6.923& 0.070 \\
&CO& 1448.3& 12.3& 6.905& 0.059 & 1447.8& 13.3& 6.907& 0.063 & -  & -  & -  & -\\

&CO$_2$& 1446.8& 9.9& 6.912& 0.047 &   1446.4& 8.1& 6.914& 0.039 & -  & -  & -  & -\\

&CH$_4$ & 1446.8& 12.2& 6.912& 0.058& 1446.4& 11.1& 6.914& 0.053 & 1445.9& 11.0& 6.916& 0.053 \\

&NH$_3$ & 1446.8& 19.7& 6.912& 0.0942 & 1447.3& 19.7& 6.909& 0.094  & 1445.9& 18.4& 6.916& 0.088 \\
&H$_2$O:CO$_2$ & -  & -  & -  & -  & 1446.4& 13.7& 6.914& 0.066 & -  & -  & -  & -\\

& 4 comp&   & -  & -  & -  & -  & -  & -  & -  & 1445.9& 14.8& 6.916& 0.071  \\\hline   
\multirow{5}{*}{50} &Pure ice & 1448.0& 18.8& 6.906& 0.089 & - &-  &  -& - & - & - & - &- \\

&H$_2$O & 1445.4& 15.2& 6.9186& 0.072  & -  & -  & -  & -  & 1444.4& 14.3& 6.923& 0.069 \\


&CO$_2$ & 1446.4& 10.3& 6.914& 0.049  &   1446.4& 8.2& 6.914& 0.039 & -  & -  & -  & -\\

&CH$_4$ & 1447.8& 16.8& 6.907& 0.0780 & 1446.8& 14.9& 6.911& 0.071  & 1445.9& 16.8& 6.916& 0.080 \\

&NH$_3$ & 1446.8& 19.0& 6.912& 0.0908 & 1445.9& 18.2& 6.916& 0.087  & 1445.4& 18.1& 6.919& 0.087 \\
&H$_2$O:CO$_2$ & -  & -  & -  & -  & 1445.9& 13.5& 6.9163& 0.0648  & -  & -  & -  & -\\

& 4 comp&   & -  & -  & -  & -  & -  & -  & -  & 1444.4& 14.5& 6.923& 0.069  \\ \hline
\multirow{5}{*}{80} &Pure ice & 1447.6& 18.5& 6.908& 0.088  & - &-  &  -& - & - & - & - &- \\

&H$_2$O & 1444.9& 14.9& 6.921& 0.071  & -  & -  & -  & -  & 1443.9& 13.5& 6.925& 0.065 \\

&CO$_2$ & 1446.4& 11.9& 6.914& 0.057  & 1454.5& 12.8& 6.875& 0.060  & -  & -  & -  & -\\

&NH$_3$ & 1446.8& 18.3& 6.912& 0.0874 & 1445.9& 17.9& 6.916& 0.085  & 1445.4& 17.4& 6.919& 0.083 \\
&H$_2$O:CO$_2$ & -  & -  & -  & -  & 1445.9& 13.7& 6.916& 0.066 & -  & -  & -  & -\\

& 4 comp&   & -  & -  & -  & -  & -  & -  & -  & 1443.5& 14.8& 6.928& 0.071  \\\hline

\multirow{5}{*}{100} &Pure ice& 1454.1& 12.1& 6.877& 0.057 & - &-  &  -& - & - & - & - &- \\

&H$_2$O & 1444.9& 15.0& 6.921& 0.072 & -  & -  & -  & -  & 1443.9& 13.5& 6.925& 0.065 \\

&CO$_2$& 1454.1& 15.8& 6.877& 0.075  & -  & -  & -  & -  & -  & -  & -  & -\\

&NH$_3$ & 1446.8& 21.0& 6.912& 0.100& 1454.1& 21.9& 6.877& 0.103  & 1445.9& 20.6& 6.916& 0.099 \\
&H$_2$O:CO$_2$ & -  & -  & -  & -  & 1445.9& 14.6& 6.916& 0.070 & -  & -  & -  & -\\

& 4 comp&   & -  & -  & -  & -  & -  & -  & -  & 1443.9& 15.3& 6.925& 0.073  \\\hline

\multirow{5}{*}{120} &Pure ice & 1453.6& 5.7& 6.880& 0.027  & - &-  &  -& - & - & - & - &- \\

&H$_2$O & 1445.4& 16.6& 6.919& 0.079 & -  & -  & -  & -  & 1443.9& 13.8& 6.925& 0.066 \\
&NH$_3$& 1454.1& 6.0& 6.8773& 0.028& 1454.1& 5.5& 6.877& 0.026 & 1445.9& 20.6& 6.916& 0.099 \\
&H$_2$O:CO$_2$ & -  & -  & -  & -  & 1445.9& 16.8& 6.916& 0.080 & -  & -  & -  & -\\

& 4 comp& -  & -  & -  & -  &   & -  & -  & -  & 1443.9& 15.7& 6.925& 0.075 \\\hline

\multirow{5}{*}{140} &Pure ice& 1453.1& 6.3& 6.882& 0.030 & - &-  &  -& - & - & - & - &- \\

&H$_2$O & 1453.1& 7.8& 6.882& 0.037  & -  & -  & -  & -  & 1452.6& 11.4& 6.884& 0.054 \\
&H$_2$O:CO$_2$ & -  & -  & -  & -  & 1453.1& 8.0& 6.882& 0.038 & -  & -  & -  & -\\

& 4 comp& -  & -  & -  & -  & -  & -  & -  & -  & 1453.1& 7.4& 6.882& 0.035 \\
\hline    
\end{tabular}
\centering
  \end{threeparttable}
  \label{1450-mixtures}
\end{table*}

\begin{table*}[h]
\centering
\setlength{\tabcolsep}{3.5pt} 
\renewcommand{\arraystretch}{1.1} 

\fontsize{9}{10}\selectfont 
\caption[]{Peak position and FWHM of methyl cyanide CH$_3$ antisymmetric deformation mode (1410 cm$^{-1}$/ 7.092 $\mu$m) in pure and mixed ices at temperatures ranging from 15  - 150 K.}       
\begin{threeparttable}[b]
\begin{tabular}{l|l|cc|cc|cc|cc|cc|cc}     
\hline           
\multirow{2}{*}{T} & \multirow{2}{*}{Matrix}&  \multicolumn{4}{c|}{1:5}& \multicolumn{4}{c|}{1:10}& \multicolumn{4}{c}{1:20}  \\

\cline{3-14}
 & -  & \multirow{2}{*}{Peak}& \multirow{2}{*}{FWHM} & \multirow{2}{*}{Peak}& \multirow{2}{*}{FWHM} &\multirow{2}{*}{Peak}& \multirow{2}{*}{FWHM} &\multirow{2}{*}{Peak}& \multirow{2}{*}{FWHM}&\multirow{2}{*}{Peak}& \multirow{2}{*}{FWHM}&\multirow{2}{*}{Peak}& \multirow{2}{*}{FWHM} \\
& -  & -  & -  & & -  & & -  & &&& -  & & -\\ 
& -  & (cm$^{-1}$) &(cm$^{-1}$)  & ($\mu$m)& ($\mu$m) &(cm$^{-1}$) &(cm$^{-1}$)  & ($\mu$m)& ($\mu$m)&(cm$^{-1}$) &(cm$^{-1}$)  & ($\mu$m)& ($\mu$m)\\ 
 
\hline  

\multirow{8}{*}{15}&Pure ice & 1410.0& 21.0& 7.092& 0.106& -  & -  & -  & -  & -  & -  & -\\
&H$_2$O & 1411.6& 19.0& 7.084& 0.095& 1411.2& 18.3& 7.086& 0.092  & 1411.2& 18.9& 7.086& 0.095 \\
&CO& 1411.6& 12.1& 7.084& 0.061 & 1410.2& 12.0& 7.091& 0.060  & -  & -  & -  & -\\

&CO$_2$& 1411.6& 9.2& 7.084& 0.046 & 1411.6& 7.5& 7.084& 0.038 & -  & -  & -  & -\\
&CH$_4$ & 1411.2& 12.6& 7.086& 0.063 & 1410.2& 11.7& 7.091& 0.059& 1408.3& 14.5& 7.101& 0.073\\

&NH$_3$ & 1414.0& 26.3& 7.072& 0.131 & 1415.0& 26.7& 7.067& 0.133 & 1412.6& 21.1& 7.079& 0.106 \\
&H$_2$O:CO$_2$ & -  & -  & -  & -  & 1412.1& 15.1& 7.082& 0.076 & -  & -  & -  & -\\
& 4 comp& -  & -  & -  & -  & -  & -  & -  & -  & 1409.7& 20.1& 7.093& 0.101 \\\hline   

\multirow{8}{*}{30}&Pure ice & 1410.0& 21.0& 7.092& 0.105  & -  & -  & -  & -  & -  & -  & -  & -\\
 &H$_2$O & 1410.7& 18.4& 7.089& 0.093 & 1410.2& 18.0& 7.091& 0.090  & 1411.2& 18.3& 7.086& 0.092 \\
&CO& 1411.2& 19.2& 7.086& 0.096& 1412.6& 16.8& 7.079& 0.0084  & -  & -  & -  & -\\

&CO$_2$ & 1411.6& 9.4& 7.084& 0.047 & 1411.6& 7.7& 7.084& 0.039  & -  & -  & -  & -\\
&CH$_4$ & 1411.6& 12.8& 7.084& 0.064  & 1411.2& 11.2& 7.086& 0.056  & -  & -  & -  & -\\

&NH$_3$ & 1412.1& 25.7& 7.082& 0.129 & 1413.6& 26.7& 7.074& 0.134 & 1412.6& 20.0& 7.079& 0.100 \\
&H$_2$O:CO$_2$ & -  & -  & -  & -  & 1411.6& 15.1& 7.084& 0.076 & -  & -  & -  & -\\

& 4 comp& -  & -  & -  & -  & -  & -  & -  & -  & 1411.6& 19.6& 7.084& 0.098\\\hline   

\multirow{8}{*}{50}&Pure ice & 1410.2& 21.1& 7.091& 0.106  & -  & -  & -  & -  & -  & -  & -  & -\\
 &H$_2$O& 1411.2& 18.3& 7.086& 0.092& 1410.2& 18.5& 7.091& 0.093 & 1411.2& 18.2& 7.086& 0.092 \\
&CO& 1413.1& 24.3& 7.077& 0.122  & 1414.0& 24.2& 7.077& 0.121 & -  & -  & -  & -\\

&CH$_4$& 1411.2& 19.0& 7.086& 0.095 & 1411.2& 17.6& 7.086& 0.088  & -  & -  & -  & -\\

&NH$_3$& 1414.0& 26.1& 7.072& 0.130 & 1414.5& 27.8& 7.069& 0.139  & 1412.6& 19.2& 7.079& 0.096 \\
&H$_2$O:CO$_2$ & -  & -  & -  & -  & 1412.1& 14.9& 7.082& 0.075 & -  & -  & -  & -\\

& 4 comp& -  & -  & -  & -  & -  & -  & -  & -  & 1411.2& 19.2& 7.086& 0.096\\ \hline
\multirow{7}{*}{80}& Pure ice &1410.9& 22.0& 7.088& 0.111 & -  & -  & -  & -  & -  & -  & -  & -\\

&H$_2$O & 1411.2& 19.2& 7.086& 0.096 & 1411.2& 18.5& 7.086& 0.093 & 1411.2& 18.1& 7.086& 0.091 \\

&CO$_2$ & 1412.6& 10.7& 7.079& 0.054  &1408.7 & -  & 7.100 & -  & -  & -  & -  & -\\
& -  & & -  & & -  & 1419.3& -  &7.045 & -  & -  & -  & -  & -\\


&NH$_3$ & 1413.1& 26.5& 7.077& 0.133 & 1413.1& 22.5& 7.077& 0.113 & 1413.1& 18.1& 7.077& 0.091 \\
&H$_2$O:CO$_2$ & -  & -  & -  & -  & 1412.1& 18.5& 7.082& 0.093 & -  & -  & -  & -\\

& 4 comp& -  & -  & -  & -  & -  & -  & -  & -  & 1410.7& 18.9& 7.089& 0.095 \\\hline

\multirow{8}{*}{100}&Pure ice & 1408.5& 4.8& 7.100& 0.024  & -  & -  & -  & -  & -  & -  & -  & -\\
&   & 1416.6 & -  & 7.060&   & -  & -  & -  & -  & -  & -  & -  & -\\
&   & 1419.8 & -  &7.043 &   & -  & -  & -  & -  & -  & -  & -  & -\\

&H$_2$O & 1411.6& 19.6& 7.084& 0.098& 1410.7& 19.2& 7.089& 0.096 & 1411.2& 18.3& 7.086& 0.092 \\
&CO$_2$ & 1416.0& 19.8& 7.062& 0.099  & -  & -  & -  & -  & -  & -  & -  & -\\

&NH$_3$ & -  & -  & -  & -  & 1416.5& - & 7.060& -  & 1412.6& 21.2& 7.079& 0.106 \\
&H$_2$O:CO$_2$ & -  & -  & -  & -  & 1412.6& 21.2& 7.079& 0.106 & -  & -  & -  & -\\

& 4 comp& -  & -  & -  & -  & -  & -  & -  & -  & 1410.7& 18.8& 7.090& 0.094\\\hline

\multirow{8}{*}{120}&Pure ice & 1408.5& 2.6& 7.100& 0.013 & -  & -  & -  & -  & -  & -  & -  & -\\
& -  &  1416.2 & -  & 7.061&   & -  & -  & -  & -  & -  & -  & -  & -\\
& -  & 1419.6  & -  &7.044 &   & -  & -  & -  & -  & -  & -  & -  & -\\
&H$_2$O & 1409.2& 23.1& 7.096& 0.116 & 1410.7& 21.1& 7.089& 0.106  & 1411.2& 20.1& 7.086& 0.101 \\

&NH$_3$ & 1408.7& 3.1& 7.098& 0.036&1408.7 & 3.1681 & 7.099  & 0.016  & 1412.6& 21.2& 7.079& 0.106 \\
&  & 1416.4*& 7.6& 7.060* & 0.038  &  1416.5*& - & 7.060 & - & -  & -  & -  & -\\
&   & 1419.8*& -  & 7.043* & -  &1419.8* & 8.016  &7.043 & 0.040 & -  & -  & -  & -\\
&H$_2$O:CO$_2$ & -  & -  & -  & -  & 1412.6& 22.7& 7.079& 0.114 & -  & -  & -  & -\\

& 4 comp& -  & -  & -  & -  & -  & -  & -  & -  & 1411.2& 18.4& 7.086& 0.092 \\\hline

\multirow{10}{*}{140}&Pure ice  & 1408.7 & 3.36 & 7.098 & 0.017 & -  & -  & -  & -  & -  & -  & -  & -\\
&   & 1416.0* &      & 7.062* & -  & & -  & -  & -  & -  & -  & -  & -\\
&   & 1419.1* & 7.1& 7.047* & 0.035 && -  & -  & -  & -  & -  & -  & -\\

 &H$_2$O & 1408.7& 7.1& 7.098* & 0.022 & 1408.7& 4.9& 7.098& 0.025  & 1408.7 & 4.6& 7.098 & 0.023\\
 &      & 1416.0*& 8.9 & 7.060*  & 0.044   &1416.0 & - & 7.062&     & -  & -  & -  & -\\
&       & 1418.9*& -  &    7.043 &1418.9  && -  & -  &   &1418.4 &9.6 &7.050 &0.048 \\

&H$_2$O:CO$_2$ & -  & -  & -  & -  & 1408.7& 4.704& 7.098& 0.024 & -  & -  & -  & -\\
&   & & -  & -  & -  & 1416.0* &  8.93 & 7.062 & 0.044 &   & -  & -  & -\\
&   & & -  & -  & -  & 1418.8* &   & - & 7.048 &       & -  & -  & -\\

& 4 comp& -  & -  & -  & -  & -  & -  & -  & -  & 1409.2& 5.9& 7.0961& 0.0297 \\\hline

\multirow{2}{*}{150}& Pure & 1409.0& 4.0& 7.097& 0.0201 & -  & -  & -  & -  & -  & -  & -\\
&   & 1419.1& 7.6& 7.046& 0.038 &     & -  & -  & -  & -  & -  & -  & -\\
\hline    
\end{tabular}
\centering
   
  \end{threeparttable}
  \label{1400-mixtures}
\end{table*}

\begin{table*}[h]
\centering
\setlength{\tabcolsep}{3.5pt} 
\renewcommand{\arraystretch}{1.1} 

\fontsize{9}{10}\selectfont 
\caption[]{Peak position and FWHM of methyl cyanide CH$_3$ symmetric deformation mode (1374.5 cm$^{-1}$/ 7.275 $\mu$m) in pure and mixed ices at temperatures ranging from 15 - 150 K.}       
\begin{threeparttable}[b]
\begin{tabular}{l|l|cc|cc|cc|cc|cc|cc}     
\hline           
\multirow{2}{*}{T} & \multirow{2}{*}{Matrix}&  \multicolumn{4}{c|}{1:5}& \multicolumn{4}{c|}{1:10}& \multicolumn{4}{c}{1:20}  \\

\cline{3-14}
 &   & \multirow{2}{*}{Peak}& \multirow{2}{*}{FWHM} & \multirow{2}{*}{Peak}& \multirow{2}{*}{FWHM} &\multirow{2}{*}{Peak}& \multirow{2}{*}{FWHM} &\multirow{2}{*}{Peak}& \multirow{2}{*}{FWHM}&\multirow{2}{*}{Peak}& \multirow{2}{*}{FWHM}&\multirow{2}{*}{Peak}& \multirow{2}{*}{FWHM} \\
&   &   &   & &   & &   & &&&   & & \\ 
&   & (cm$^{-1}$) &(cm$^{-1}$)  & ($\mu$m)& ($\mu$m) &(cm$^{-1}$) &(cm$^{-1}$)  & ($\mu$m)& ($\mu$m)&(cm$^{-1}$) &(cm$^{-1}$)  & ($\mu$m)& ($\mu$m)\\ 
 
\hline    
\multirow{4}{*}{15} &Pure& 1374.5& 15.5& 7.275& 0.0821  & -  & -  & -  & -  & -  & -  & -  & -\\

&H$_2$O& 1374.5& 13.5& 7.275& 0.072 & 1374.5& 13.4& 7.275& 0.071  & 1374.5& 13.9& 7.275& 0.074 \\
&CO& 1377.4& 7.9& 7.260& 0.041  & 1376.4& 7.4& 7.265& 0.039  & -  & -  & -  & -\\

&CO$_2$ & 1375.5& 7.7& 7.270& 0.041  & 1375.0& 6.7& 7.273& 0.036  & -  & -  & -  & -\\
&CH$_4$ & 1377.4& 7.4& 7.260& 0.039 & 1376.9& 6.9& 7.263& 0.036 & 1375.5& 7.9& 7.270& 0.042\\

&NH$_3$ & 1375.0& 14.9& 7.273& 0.079 & 1375.0& 15.5& 7.273& 0.082  & 1374.5& 16.2& 7.275& 0.086 \\
&H$_2$O:CO$_2$ & -  & -  & -  & -  & 1374.5& 12.3& 7.275& 0.065 & -  & -  & -  & -\\

& 4 comp& -  & -  & -  & -  & -  & -  & -  & -  & 1376.0& 14.0& 7.268& 0.0740\\\hline    
\multirow{4}{*}{30}  &Pure& 1374.3& 15.3& 7.277& 0.081& -  & -  & -  & -  & -  & -  & -  & -\\

&H$_2$O & 1375.0& 13.1& 7.273& 0.069 & 1375.0& 13.0& 7.273& 0.069  & 1373.1& 14.5& 7.283& 0.077\\
&CO& 1377.9& 8.8& 7.257& 0.047  & 1377.9& 9.8& 7.257& 0.052  & -  & -  & -  & -\\

&CO$_2$ & 1375.5& 7.5& 7.270& 0.040 & 1375.0& 6.6& 7.273& 0.035 & -  & -  & -  & -\\
&CH$_4$ & 1377.4& 7.5& 7.260& 0.040  & 1377.4& 6.7& 7.260& 0.035 & 1376.9& 7.6& 7.263& 0.040 \\

&NH$_3$ & 1375.5& 14.7& 7.270& 0.078 & 1375.0& 15.4& 7.273& 0.081  & 1373.1& 16.0& 7.283& 0.085 \\
&H$_2$O:CO$_2$ & -  & -  & -  & -  & 1375.0& 11.9& 7.273& 0.063  & -  & -  & -  & -\\

& 4 comp& -  & -  & -  & -  & -  & -  & -  & -  & 1375.0& 14.6& 7.273& 0.077\\\hline   
\multirow{4}{*}{50} &Pure& 1374.5& 14.8& 7.275& 0.078  & -  & -  & -  & -  & -  & -  & -  & -\\
&H$_2$O& 1374.5& 12.1& 7.275& 0.064 & 1374.5& 12.0& 7.275& 0.063  & 1372.6& 14.0& 7.286& 0.074 \\

&CO$_2$ & 1375.5& 7.5& 7.270& 0.040  & 1375.0& 6.5& 7.273& 0.035 & -  & -  & -  & -\\
&CH$_4$ & 1375.0& 13.2& 7.273& 0.070 & 1374.0& 13.3& 7.278& 0.071  & 1373.1& 14.7& 7.283& 0.078\\

&NH$_3$ & 1375.5& 14.3& 7.270& 0.075& 1375.0& 15.3& 7.273& 0.081 & -  & -  & -  & -\\
&H$_2$O:CO$_2$ & -  & -  & -  & -  & 1374.5& 10.8& 7.275& 0.057  & -  & -  & -  & -\\

& 4 comp& -  & -  & -  & -  & -  & -  & -  & -  & 1375.0& 12.7& 7.273& 0.067 \\ \hline
\multirow{4}{*}{80}&Pure& 1374.5& 14.2& 7.275& 0.075 & -  & -  & -  & -  & -  & -  & -  & -\\
&H$_2$O & 1373.6& 11.1& 7.280& 0.059 & 1374.5& 11.1& 7.275& 0.058  & 1373.1& 12.9& 7.283& 0.068\\

&CO$_2$ & 1375.5& 7.8& 7.270& 0.041 & 1377.9& 7.9& 7.257& 0.042  & -  & -  & -  & -\\

&NH$_3$ & 1375.0& 12.9& 7.273& 0.068 & 1375.0& 14.0& 7.273& 0.074 & 1373.6& 14.8& 7.280& 0.078 \\
&H$_2$O:CO$_2$ & -  & -  & -  & -  & 1374.5& 10.0& 7.275& 0.053  & -  & -  & -  & -\\

& 4 comp& -  & -  & -  & -  & -  & -  & -  & -  & 1375.0& 12.6& 7.273& 0.066 \\\hline

\multirow{4}{*}{100} &Pure& -  & -  & -  & -  & -  & -  & -  & -  & -  & -  & -  & -\\
&H$_2$O & 1374.0& 10.3& 7.278& 0.054 & 1374.5& 10.3& 7.275& 0.054 & 1373.6& 12.5& 7.280& 0.066 \\

&CO$_2$& 1377.9& 10.7& 7.257& 0.056 & -  & -  & -  & -  & -  & -  & -  & -\\

&NH$_3$ & 1377.9& 10.3& 7.257& 0.054 & 1377.9& 11.9& 7.257& 0.062 & 1374.0& 13.1& 7.278& 0.070 \\
&H$_2$O:CO$_2$ & -  & -  & -  & -  & 1375.0& 9.6& 7.273& 0.051  & -  & -  & -  & -\\

& 4 comp& -  & -  & -  & -  & -  & -  & -  & -  & 1375.5& 12.5& 7.270& 0.066 \\\hline

\multirow{4}{*}{120}&Pure& -  & -  & -  & -  & -  & -  & -  & -  & -  & -  & -  & -\\
&H$_2$O & 1374.0& 10.1& 7.278& 0.054 & 1374.0& 10.3& 7.278& 0.055 & 1373.6& 11.1& 7.280& 0.059 \\

&NH$_3$ &  1371.6  & 2.4 & 7.291 &0.013 & 1377.9& 2.4& 7.257& 0.013  & 1374.0& 13.1& 7.278& 0.070 \\
& -  & 1377.9   & 2.4 &7.257  &0.013  & -  & -  & -  & -  & -  & -  & -  & -\\
&H$_2$O:CO$_2$ & -  & -  & -  & -  & 1374.5& 10.0& 7.275& 0.053  & -  & -  & -  & -\\

& 4 comp& -  & -  & -  & -  & -  & -  & -  & -  & 1375.0& 10.8& 7.27& 0.057 \\\hline

\multirow{4}{*}{140} &Pure& -  & -  & -  & -  & -  & -  & -  & -  & -  & -  & -  & -\\
&H$_2$O &  1371.6  & 3.1 & 7.291 & 0.016 & 1377.4& 2.8& 7.260& 0.015  & -  & -  & -  & -\\
&       &   1377.4 & 2.6 & 7.260 & 0.014 & -  & -  & -  & -  & -  & -  & -  & -\\
&H$_2$O:CO$_2$ & -  & -  & -  & -  &1371.6  &2.9  & 7.291 & 0.014 & -  & -  & -  & -\\

&              & -  & -  & -  & -  &1377.4  & 2.6  & 7.260 & 0.015 & -  & -  & -  & -\\
& 4 comp& -  & -  & -  & -  & -  & -  & -  & -  & 1377.4& 3.0& 7.260& 0.016 \\
\hline    
\end{tabular}
\centering
  \end{threeparttable}
  \label{1370-mixtures}
\end{table*}

\begin{table*}[h]
\centering
\setlength{\tabcolsep}{3.5pt} 
\renewcommand{\arraystretch}{1.1} 

\fontsize{9}{10}\selectfont
\caption[]{Peak position and FWHM of methyl cyanide CH$_3$ rock (1041.6 cm$^{-1}$/ 9.600 $\mu$m) in pure and mixed ices at temperatures ranging from 15  - 150 K.}       
\begin{threeparttable}[b]
\begin{tabular}{l|l|cc|cc|cc|cc|cc|cc}     
\hline           
\multirow{2}{*}{T} & \multirow{2}{*}{Matrix}&  \multicolumn{4}{c|}{1:5}& \multicolumn{4}{c|}{1:10}& \multicolumn{4}{c}{1:20}  \\

\cline{3-14}
 &   & \multirow{2}{*}{Peak}& \multirow{2}{*}{FWHM} & \multirow{2}{*}{Peak}& \multirow{2}{*}{FWHM} &\multirow{2}{*}{Peak}& \multirow{2}{*}{FWHM} &\multirow{2}{*}{Peak}& \multirow{2}{*}{FWHM}&\multirow{2}{*}{Peak}& \multirow{2}{*}{FWHM}&\multirow{2}{*}{Peak}& \multirow{2}{*}{FWHM} \\
&   &   &   & &   & &   & &&&   & & \\ 
&   & (cm$^{-1}$) &(cm$^{-1}$)  & ($\mu$m)& ($\mu$m) &(cm$^{-1}$) &(cm$^{-1}$)  & ($\mu$m)& ($\mu$m)&(cm$^{-1}$) &(cm$^{-1}$)  & ($\mu$m)& ($\mu$m)\\ 
 
\hline    
\multirow{5}{*}{15} &Pure ice&1041.6& 19.9& 9.600& 0.183 & -  & & -  & -  &  - & - & - &- \\
&H$_2$O& 1040.9& 14.6& 9.607& 0.1351  & 1041.4& 14.0& 9.603& 0.129 & 1041.4& 13.7& 9.603& 0.126 \\
&CO  & 1040.9& 8.6& 9.607& 0.079  & 1040.4& 9.0& 9.612& 0.083 & -  & -  & -  & -\\

&CO$_2$& 1039.9& 11.1& 9.616& 0.103 & 1039.0& 10.6& 9.625& 0.098 & -  & -  & -  & -\\

&CH$_4$& 1041.9& 7.7& 9.598& 0.071  & 1040.9& 6.9& 9.607& 0.063 & 1040.4& 6.3& 9.612& 0.058 \\

&NH$_3$ & 1037.5& 13.4& 9.638& 0.124  & 1037.5& 12.3& 9.638& 0.115& 1037.0& 11.4& 9.643& 0.106\\
&H$_2$O:CO$_2$ & -  & -  & -  & -  & 1039.9& 14.4& 9.612& 0.133 & -  & -  & -  & -\\

& 4 comp& -  & -  & -  & -  & -  & -  & -  & -  & 1040.9& 12.6& 9.607& 0.116 \\\hline    
\multirow{5}{*}{30} &Pure ice& 1040.9& 19.7& 9.607& 0.182 && -  & & -  &  - & - & - &- \\
&H$_2$O& 1040.9& 13.8& 9.607& 0.127  & 1041.4& 12.8& 9.603& 0.118 & 1040.4& 13.7& 9.612& 0.127 \\
&CO  & 1041.4& 9.7& 9.603& 0.089  & 1040.4& 14.3& 9.612& 0.132  & -  & -  & -  & -\\

&CO$_2$& 1039.9& 11.0& 9.616& 0.102 & 1039.9& 10.6& 9.616& 0.098  & -  & -  & -  & -\\

&CH$_4$ & 1042.3& 7.8& 9.594& 0.072 & 1041.9& 6.9& 9.598& 0.064 & 1041.9& 6.1& 9.598& 0.057 \\

&NH$_3$ & 1037.5& 13.5& 9.638& 0.126 & 1037.5& 12.8& 9.638& 0.119 & 1037.0& 11.0& 9.643& 0.102 \\
&H$_2$O:CO$_2$ & -  & -  & -  & -  & 1039.4& 14.0& 9.616& 0.129 & -  & -  & -  & -\\

& 4 comp& -  & -  & -  & -  & -  & -  & -  & -  & 1040.9& 11.3& 9.607& 0.104 \\\hline   
\multirow{5}{*}{50} &Pure ice& 1041.4& 18.7& 9.603& 0.173&& -  & & -  &  - & - & - &- \\

&H$_2$O & 1040.4& 13.5& 9.612& 0.125 & 1040.4& 12.1& 9.612& 0.112  & 1039.9& 14.0& 9.616& 0.130 \\

&CO$_2$& 1039.9& 10.8& 9.616& 0.100 & 1039.9& 10.3& 9.616& 0.095 & -  & -  & -  & -\\

&CH$_4$& 1041.9& 15.7& 9.598& 0.145 & 1040.9& 15.8& 9.607& 0.146 & 1041.9& 14.8& 9.598& 0.136 \\

&NH$_3$ & 1037.5& 12.8& 9.638& 0.119 & 1037.0& 12.8& 9.643& 0.119  & -  & -  & -  & -\\
&H$_2$O:CO$_2$ & -  & -  & -  && 1039.4& 13.6& 9.629& 0.126 & -  & -  & -  & -\\

& 4 comp& -  & -  & -  & -  & -  & -  & -  & -  & 1040.4& 10.7& 9.612& 0.099 \\ \hline
\multirow{6}{*}{80} &Pure ice&  1041.1& 17.5& 9.605& 0.162& -  & &   & - & - & - &- \\
&H$_2$O & 1040.4& 12.4& 9.612& 0.115 & 1039.9& 11.3& 9.616& 0.104 & 1039.0& 14.1& 9.625& 0.130 \\

&CO$_2$& 1039.9& 11.6& 9.616& 0.107 & 1036.1& 9.2& 9.652& 0.086  & -  & -  & -  & -\\


&NH$_3$ & 1037.5& 12.6& 9.638& 0.117 & 1037.0& 12.5& 9.643& 0.117 & -  & -  & -  & -\\
&H$_2$O:CO$_2$ & -  & -  & -  & -  & 1039.0& 12.8& 9.621& 0.119 & -  & -  & -  & -\\

& 4 comp& -  & -  & -  & -  & -  & -  & -  & -  & 1040.4& 10.4& 9.612& 0.096 \\\hline

\multirow{8}{*}{100} &Pure ice& 1036.1*& -  &  9.652& 0.102  & -  & & -  & -  & -  & -  &   \\
&   &1040.2*& 15.07 &9.613    & 0.139  & -  & -  & & -  & -  & -  & -  &   \\
&   &1048.4*& -  &9.539 & 0.037 & -  & -  & & -  & -  & -  & -  &   \\

&H$_2$O & 1039.4& 12.2& 9.621& 0.113  & 1040.4& 11.7& 9.612& 0.108  & 1038.5& 13.9& 9.629& 0.129 \\

&CO$_2$ & 1040.4& 14.8& 9.612& 0.137  & -  & -  & -  & -  & -  & -  & -  & -\\

&NH$_3$ & 1039.9& 13.1& 9.616& 0.121  & -  & -  & -  & -  & -  & -  & -  & -\\
&H$_2$O:CO$_2$ & -  & -  & -  & -  & 1039.0& 13.1& 9.621& 0.121 & -  & -  & -  & -\\

& 4 comp& -  & -  & -  & -  & -  & -  & -  & & 1039.9& 9.9& 9.616& 0.091\\\hline
\multirow{7}{*}{120} &Pure ice& 1036.1&2.11 &9.652 & 0.020  & - &-  &  -& - & - & - & - &- \\
&  & 1039.9 & 2.45&9.616 & 0.022 & - &-  &  -& - & - & - & - &- \\
&   & 1048.1& 2.35& 9.541& 0.021 & - &-  &  -& - & - & - & - &- \\

&H$_2$O & 1039.9& 12.3& 9.616& 0.114 & 1040.4& 11.8& 9.612& 0.109  & 1035.6& 12.9& 9.656& 0.120\\

&NH$_3$ &   1036.1* &  7.3 &  9.652* & 0.0684 & -  & -  & -  & -  & -  & -  & -  & -\\
&       &   1040.0* & -  & 9.616* & -  & -  & -  & -  & -  & -  & -  & -  & -\\
&       &  1048.1  & 3.02 & 9.541 & 0.027 & -  & -  & -  & -  & -  & -  & -  & -\\

&H$_2$O:CO$_2$ & -  & -  & -  & -  & 1038.5& 14.4& 9.621& 0.134  & -  & -  & -  & -\\

& 4 comp& -  & -  & -  & -  & -  & -  & -  & -  & 1040.9& 8.2& 9.607& 0.075\\\hline

\multirow{7}{*}{140} &Pure ice&1036.3 & 2.3 &9.650 & 0.022 & - &-  &  -& - & - & - & -  \\
&   & 1039.9& 2.7&  9.616 & 0.025  & -  & -  & & -  & -  &   & -  & -\\
&   & 1047.6& 2.54& 9.545&0.023 & -  & -  & & -  & -  & -  & -  &   \\

&H$_2$O  &1036.1* & 6.3& 9.652 &0.059 & 1036.1*&7.3 &9.652* & 0.068 & -  & -  & -  & -\\
&   & 1040.0* & -  & & -  &1039.9* & -  &9.616* & -  & -  & -  & -  & -\\
&   & 1047.6 & 3.2 & 9.545& 0.029& 1047.6& 3.6& 9.545& 0.033 & -  & -  & -  & -\\

&H$_2$O:CO$_2$ & -  & -  & -  & -  &1036.1  & 7.0 & 9.652 & 0.065 & -  & -  & -  & -\\
&   & -  & -  & -  & -  &1047.6  &3.5  & 9.545 & 0.032 & -  & -  & -  & -\\

& 4 comp& -  & -  & -  & -  & -  & -  & -  & -  & -  & -  & -  & -\\
\hline
\multirow{1}{*}{150} &Pure ice& 1038.0& 6.9& 9.634& 0.064 & - &-  &  -& - & - & - & - &- \\
&   &1039.7 & - & 9.618& - & -  & -  & -  & -  & -  & -  & -  &   \\
&   &1047.4 &2.8 & 9.547 & 0.025  & -  & -  & -  & -  &   & -  & -  & -\\
\hline    
\end{tabular}
\centering
  \end{threeparttable}
  \label{1040-mixtures}
\end{table*}

%% file: AppendixC-areas.tex
This section presents the integrated absorbance of methyl cyanide bands (i.e., band areas) in ice mixtures at different temperatures. Each table lists the integrated absorbance of the absorption bands in a specific ice mixture normalized to the integrated absorbance of the CN stretch in that mixture at 15 K. These values are used to calculate the apparent band strength of the methyl cyanide features in ice mixtures at different temperatures, as explained in the following.

The apparent band strength of the CN stretch feature of methyl cyanide in the IR spectrum of an ice mixture (A$_{CN, mixt}^{'}$ (T)) is related to the band strength of another feature, \emph{i}, (A$_{i, mixt}^{'}$ (T)) by:

\begin{equation} A_{i, mixt}^{'} (T)= \frac{a_{i, mixt}(T)}{a_{CN, mixt} (T)} \times A_{CN, mixt}^{'} (T),\label{bs1}\end{equation}
where  a$_{CN, mixt}$ and a$_{i, mixt}$ are the integrated absorbances of the CN stretching and the feature \emph{i}, respectively. This relation can be rewritten in terms of the band strengths of the pure methyl cyanide at 15 K and the relative band strength of the CN stretching in the ice mixture (both derived in this work):

\begin{equation}
    A_{i, mixt}^{'} (T) =   \frac{a_{i,mixt}(T)}{a_{CN, mixt} (15 K)} \times \eta_{CN, mixt} \times A_{CN} (15 K) ,
    \label{relativebs_final}
\end{equation}
where $\eta_{CN,mixt}$ is the relative band strength of the CN stretch in the analyzed ice mixture (Appendix A), A$_{CN}$ (15 K) is the band strength for the pure CN stretch mode at 15 K (Table \ref{band-strength}), and the term $\frac{a_i(T)}{a_{CN, mixt}(15 K)}$ is the integrated absorbance for the band \emph{i}, in the ice mixture, normalized to the integrated absorbance of the CN stretch in the same ice mixture at 15 K (Table \ref{rel_areas}).

To illustrate the use of the data in this Section for calculating the apparent band strength in mixtures at temperatures above 15 K, we derived below the apparent band strength of the CH$_3$ rock feature of methyl cyanide in the CH$_3$CN:H$_2$O (1:10) ice at 80 K (A$_{CH_3,rock}'$ (80 K)). Using  $\eta$ =   1.75 (relative band strength value for the CN stretching in the CH$_3$CN:H$_2$O (1:10) ice mixture, shown on Figure \ref{fig:2252}), A$_{CN}$ (15 K) = 1.9 $\times$ 10$^{-18}$ cm molecule$^{-1}$ (Table \ref{band-strength}) and  $\frac{a_{CH_3,rock}(80 K)}{a_{CN} (15 K)}$ = 0.31 (Table \ref{rel_areas}) on Equation \ref{relativebs_final}, we obtain A$_{CH_3, rock}'$  = 0.31 $\times$ 1.75 $\times$ 1.9 $\times$ 10$^{-18}$ cm molecule$^{-1}$=  1.0 $\times$ 10$^{-18}$ cm molecule$^{-1}$.

\begin{longtable}{|c|c|c|c|c|c|c|}   

\caption{Integrated absorbance of CH$_3$CN features in different ice mixtures at different temperatures. The values are normalized with respect to the CN stretching integrated absorbance at 15 K in the mixture. } \label{tab:long} \\

\hline         
 & CH$_3$ rock & CH$_3$ symm. def & CH$_3$ antisymm. def. & Comb. & CN str. & CH$_3$ symm. str. \\ 

T (K) & 1041.6 cm$^{-1}$ & 1374.5 cm$^{-1}$&  1410 cm$^{-1}$& 1448.3 cm$^{-1}$& 2252.2 cm$^{-1}$ & 2940.9 cm$^{-1}$\\ 

& (9.600 $\mu$m) & (7.275 $\mu$m)& (7.092 $\mu$m) & (6.904 $\mu$m) & (4.4401 $\mu$m) & (3.400 $\mu$m)\\

\hline
\multicolumn{7}{|c|}{Pure CH$_3$CN}\\
\hline                       
15	&	0.80	&	0.65	&	1.06	&	1.44	&	1.00	&	0.29	\\
30	&	0.80	&	0.64	&	1.04	&	1.41	&	1.00	&	0.27	\\
50	&	0.77	&	0.61	&	0.98	&	1.34	&	1.01	&	0.26	\\
80	&	0.73	&	0.58	&	0.89	&	1.23	&	1.00	&	0.23	\\
100	&	 -	&	 -	&	0.94	&	1.23	&	1.05	&	0.30	\\
120	&	 -	&	 -	&	1.16	&	1.26	&	1.01	&	0.40	\\
140	&	 -	&	 -	&	1.11	&	1.18	&	0.98	&	0.40	\\
150	&	 -	&	 -	&	0.76	&	0.84	&	0.87	&	0.33	\\

\hline
\noalign{\penalty-5000}
\hline

\multicolumn{7}{|c|}{Ice mixture: CH$_3$CN:H$_2$O (1:10)}\\
\hline                       
15	&	0.42	&	0.36	&	0.70	&	1.18	&	1.00	&	0.38	\\
30	&	0.41	&	0.34	&	0.68	&	1.14	&	0.99	&	0.35	\\
50	&	0.36	&	0.33	&	0.63	&	1.05	&	0.93	&	0.31	\\
80	&	0.31	&	0.33	&	0.58	&	0.94	&	0.89	&	0.23	\\
100	&	 -	&	0.30	&	0.51	&	0.86	&	0.84	&	0.20	\\
120	&	 -	&	0.32	&	0.48	&	0.78	&	0.85	&	0.18	\\
140	&	 -	&	 -	&	 -	&	0.69	&	0.89	&	0.26	\\			\hline

\multicolumn{7}{|c|}{Ice mixture: CH$_3$CN:CO (1:10)}\\
\hline                       
15	&	1.14	&	0.49	&	0.73	&	1.31	&	1.00	&	0.54	\\
30	&	1.19	&	0.58	&	0.98	&	1.44	&	1.08	&	0.64	\\

\hline\multicolumn{7}{|c|}{Ice mixture: CH$_3$CN:CO$_2$ (1:10)}\\
\hline                       
15	&	0.64	&	0.48	&	0.58	&	0.83	&	1.00	&	0.23	\\
30	&	0.70	&	0.47	&	0.60	&	0.86	&	1.00	&	0.19	\\
50	&	0.66	&	0.49	&	0.59	&	0.85	&	1.01	&	-	\\
80 	&		-&	0.25	&	0.38	&	-	&	0.58	&	-	\\
100	&	-	&	-	&	 - 	&		-&	0.57	&	-	\\

\hline

\multicolumn{7}{|c|}{Ice mixture: CH$_3$CN:CH$_4$ (1:10)}\\
\hline                       
15	&	0.79	&	0.51	&	0.66	&	1.54	&	1.00	&	0.30	\\
30	&	0.84	&	0.54	&	0.69	&	1.45	&	1.08	&	0.35	\\
50	&	0.67	&	0.48	&	0.72	&	1.09	&	0.86	&	0.28	\\

\hline

\hline
\noalign{\penalty-5000}
\hline

\multicolumn{7}{|c|}{Ice mixture: CH$_3$CN:NH$_3$ (1:10)}\\
\hline                       
15	&	0.50	&	0.30	&	0.60	&	0.95	&	1.00	&	0.49	\\
30	&	0.84	&	0.53	&	0.59	&	1.60	&	1.83	&	0.70	\\
50	&	0.83	&	0.52	&	0.59	&	1.54	&	1.82	&	0.68	\\
80	&	0.82	&	0.50	&	0.56	&	1.41	&	1.74	&	0.60	\\
100 	&		-&	0.53	&	0.65	&	1.34	&	1.65	&	0.56	\\
120 	&		-&		&	0.54	&	1.16	&	1.17	&	0.46	\\

\hline

\hline
\multicolumn{7}{|c|}{Ice mixture: CH$_3$CN:H$_2$O:CO$_2$}\\
\hline                       
15	&	0.57	&	0.32	&	0.62	&	0.84	&	1.00	&	0.33	\\
30	&	0.56	&	0.31	&	0.62	&	0.82	&	0.99	&	0.31	\\
50	&	0.56	&	0.31	&	0.60	&	0.79	&	0.96	&	0.29	\\
80	&	0.52	&	0.30	&	0.55	&	0.69	&	0.88	&	0.26	\\
100	&	0.50	&	0.26	&	0.48	&	0.62	&	0.72	&	0.22	\\
120	&	0.48	&	0.25	&	0.44	&	0.55	&	0.68	&	0.20	\\
140	&	-	&	-	&	0.40	&	0.47	&	0.63	&	0.22	\\
\hline

\multicolumn{7}{|c|}{Ice mixture: CH$_3$CN:H$_2$O:CH$_4$:NH$_3$}\\
\hline                       
15	&	0.46	&	0.33	&	0.67	&	1.02	&	1.00	&	0.35	\\
30	&	0.41	&	0.32	&	0.66	&	0.99	&	1.00	&	0.32	\\
50	&	0.37	&	0.26	&	0.60	&	0.92	&	0.99	&	0.28	\\
80	&	0.39	&	0.28	&	0.52	&	0.82	&	0.93	&	0.18	\\
100	&	0.27	&	0.29	&	0.53	&	0.79	&	0.93	&	0.15	\\
120	&	0.19	&	0.26	&	0.46	&	0.72	&	0.91	&	0.11	\\
\hline

\label{rel_areas}
\end{longtable}